\documentclass[aps,pre,twocolumn,showpacs,floatfix,superscriptaddress,nofootinbib]{revtex4}
\usepackage{graphicx}

\newcommand {\be} {\begin{equation}}
\newcommand {\ee} {\end{equation}}
\newcommand {\Be}{\begin{eqnarray*}}
\newcommand {\Ee} {\end{eqnarray*}}
\newcommand {\bey} {\begin{eqnarray}}
\newcommand {\eey} {\end{eqnarray}}
\newcommand{\bit}{\begin{itemize}}
\newcommand{\eit}{\end{itemize}}
\newcommand{\bfl}{\begin{flusleft}}
\newcommand{\efl}{\end{flusleft}}
\newcommand{\bfr}{\begin{flushright}}
\newcommand{\bc}{\begin{center}}
\newcommand{\ec}{\end{center}}
\newcommand{\ben}{\begin{enumerate}}
\newcommand{\een}{\end{enumerate}}

\newcommand{\comment}[1]{}

\begin{document}

\title{Dynamical response of the Hodgkin-Huxley
model in the high-input regime}

\author{Stefano Luccioli}
\email{luccioli@inoa.it} \affiliation{Istituto dei Sistemi
Complessi,
Consiglio Nazionale delle Ricerche,
via Madonna del Piano 10, I-50019 Sesto Fiorentino, Italy}
\affiliation{Istituto Nazionale di Fisica Nucleare,
Sezione di Firenze,
via Sansone, 1 - I-50019 Sesto Fiorentino, Italy}
\author{Thomas Kreuz}
\email{kreuz@inoa.it} \affiliation{Istituto dei Sistemi Complessi,
Consiglio Nazionale delle Ricerche,
via Madonna del Piano 10, I-50019 Sesto Fiorentino, Italy}
\author{Alessandro Torcini}
\email{alessandro.torcini@isc.cnr.it} \affiliation{Istituto dei Sistemi
Complessi, Consiglio Nazionale delle Ricerche,
via Madonna del Piano 10, I-50019 Sesto Fiorentino, Italy}
\affiliation{Centro Interdipartimentale per lo Studio delle
Dinamiche Complesse, via Sansone, 1 - I-50019 Sesto Fiorentino,
Italy}

%%%%%%%%%%%%%%%%%%%%%%%%%%%%%%%%%%%%%%%%%%%%%%%%%%%%%%%%%%%%%%%%%%%%%%%%%%%
\begin{abstract}
The response of the Hodgkin-Huxley neuronal model
subjected to stochastic uncorrelated spike trains originating from a large
number of inhibitory and excitatory post-synaptic potentials is
analyzed in detail. The model is examined in its three
fundamental dynamical regimes: silence, bistability
and repetitive firing. Its response is characterized in
terms of statistical indicators (interspike-interval distributions and
their first moments) as well as of dynamical
indicators (autocorrelation functions and conditional entropies).
In the silent regime, the coexistence of two different coherence resonances
is revealed: one occurs at quite low noise and is
related to the stimulation of subthreshold oscillations around the rest state;
the second one (at intermediate noise variance) is associated with the
regularization of the sequence of spikes emitted by the neuron.
Bistability in the low noise limit can be interpreted in terms of jumping
processes across barriers activated by stochastic fluctuations.
In the repetitive firing regime a maximization of incoherence is observed
at finite noise variance. Finally, the mechanisms responsible for spike triggering
in the various regimes are clearly identified.
\end{abstract}
%%%%%%%%%%%%%%%%%%%%%%%%%%%%%%%%%%%%%%%%%%%%%%%%%%%%%%%%%%%%%%%%%%%%%%%%%%

% 87.19.La - Neuroscience
% 84.35.+i Neural networks
% 05.45.-a Nonlinear dynamics and nonlinear dynamical systems
% 05.45.Xt Synchronization; coupled oscillators
% 87.10.+e General theory and mathematical aspects (biophysics)
% 07.05.Mh Neural networks, fuzzy logic, artificial intelligence
% 89.75.-k Complex systems
% 87.17.Nn Electrophysiology of nerve cells
% 87.17.Aa Theory and modeling; computer simulation (cellular process)
%05.40.Ca        Noise
%05.45.Tp        Time series analysis

\maketitle

\section{Introduction}
\label{uno}

Neuronal models represent a fundamental benchmark to
investigate the dynamical response of excitable systems under the influence
of noise. One of the main reasons justifying the interest
of neuroscientists for this subject resides in the observation
that {\it in vivo} neocortical neurons are subjected to a constant
bombardment of inhibitory and excitatory post-synaptic potentials
(EPSPs and IPSPs), somehow resembling a background noise.
As a consequence in the last decades a large number of numerical
and theoretical studies have been
devoted to the characterization of the response of simple and more
elaborated neuronal models under the influence of a large variety
of stochastic inputs~\cite{tuckwell,koch,ger}.

 Among the many proposed biophysical models the one introduced
by Hodgkin and Huxley in 1952~\cite{hh} can still be considered
as a valid framework for exploring neural excitability,
due to its relative simplicity combined with the fact that it
embodies the major features of membrane potential evolution~\cite{hh_devil}.
In particular, in order to understand the origin of the
variability observed in the distribution
of spikes emitted by cortical neurons~\cite{shadlen_newsome_1998}
the response of the Hodgkin-Huxley (HH) model has recently been
studied under the influence of additive noise~\cite{tiesinga_2000,aguera03}
or subjected to trains of post-synaptic
potentials~\cite{brown_feng_1999,hasegawa,tiesinga_2000}.

{\it Stochastic resonance} (SR)~\cite{gammaitoni}
and {\it coherence resonance} (CR)~\cite{gang,pikovsky_kurths_1997}
represent some of the most interesting phenomena observed experimentally
and numerically for excitable neuronal systems driven by noise.
While SR is related to the enhanced ability of neurons to detect weak
(periodic or aperiodic) signals when subjected to additive noise, CR
refers to the regularization of the response of the system at
an optimal noise intensity in the absence of an external signal
(for a comprehensive review see \cite{lindner_ojalvo_2004}).
For neuronal systems evidences of CR have been reported experimentally
for the cat's spinal and cortical neural ensembles \cite{manjarrez}
and theoretically for the following models: FitzHugh-Nagumo \cite{pikovsky_kurths_1997},
leaky integrate-and-fire \cite{Pakdaman01, Lindner02},
Hindmarsh-Rose \cite{Longtin97} and Morris-Lecar \cite{Gong02}.
CR has also been observed for the HH model~\cite{lee_neiman_1998,yu_wang_2001},
but these results mainly refer to additive continuous noise and to
the silent regime near the saddle-node bifurcation of limit cycles.

 Our aim is to perform a detailed analysis of the response of the HH model
subjected to many stochastic trains of EPSPs and IPSPs in its
three fundamental dynamical regimes: the silent, the bistable and
the repetitive firing ones. The neuron is studied in the so called
{\it high-input regime}~\cite{shadlen_newsome_1998}, i.e. when it
receives hundreds or thousands of post-synaptic inputs per emitted
spike. In this situation the stochastic input can simply be
characterized in terms of its average, representing the
bifurcation parameter of the model, and its variance. Most
attention is devoted to the mechanisms responsible for neuronal
firing in the different dynamical regimes and to the
characterization of CR in terms of dynamical and statistical
indicators. In particular, we show that the conditional entropies
can be employed as powerful indicators to detect coherence,
similarly to what has previously been done to characterize
stochastic resonance for a Schmitt trigger~\cite{neiman_1996}.
Sound evidences of two coexisting CRs are reported, the first one
related to subthreshold oscillations occurring at very low noise
fluctuations in absence of spiking and the second one due to the
regularization of the emitted spike trains. In the regime of
bistability the intermittent dynamics between the two stable
dynamical states is interpreted in terms of jumping processes
across activation barriers induced by noise fluctuations. Finally,
evidences of incoherence maximization in the repetitive firing
regime are presented.

The model and the various employed indicators,
namely the distribution of the interspike-interval (ISI),
the spike-triggered averages,
the autocorrelation function of the signal and of the ISIs, and
the conditional entropies
are introduced in Section \ref{due}.
The results for the silent regime are reported
in Section \ref{tre}, while the bistable and repetitive firing regimes
are examined in Section \ref{quattro}
and \ref{cinque}, respectively. The last Section contains
a summary of the results and concluding remarks.

\section{Models and Tools}
\label{due}

\subsection{The Hodgkin-Huxley Model}

The Hodgkin-Huxley (HH) model describes the dynamical
evolution of the membrane potential $V(t)$ during a voltage clamp
experiment. It can be written as
\bey
&-&C\frac{dV}{dt}=g_{Na}m^{3}h(V-E_{Na})+g_{K}n^{4}(V-E_{K})+
\label{hh}
\\
&+&g_{L}(V-E_{L}) - I(t)\quad,
\nonumber
\eey
where $I(t)$ is an external current and the evolution of the
gating variables $X=m,n,h$ is ruled by three ODE's of the form
\begin{equation}
\frac{dX}{dt}=\alpha_X(V)(1-X)-\beta_X(V)X \quad .
\label{hh_gating}
\end{equation}
The parameters entering in Eq. (\ref{hh}) are $C=1 \mu$F/cm$^2$,
$E_{Na}=50$mV, $E_K$=-77 mV, $E_L$=-54.4 mV, $g_{Na}$=120
mS/cm$^2$, $g_{K}$=36 mS/cm$^2$, and $g_L$=0.3 mS/cm$^2$. The
expressions of the nonlinear functions $\alpha_X(V)$ and
$\beta_X(V)$ are explicitly reported in Table \ref{funz_HH}.

\begin{table}
\begin{center}
\begin{tabular}{c|c|c} \hline\hline
X        &  $\alpha_{X}(V)$ $(s^{-1})$               & $\beta_{X}(V)$ $(s^{-1})$ \\ \hline
m        &  0.1 (V+40)/(1-$\exp$[-(V+40)/10])        &  4 $\exp$[-(V+65)/18]  \\
n        &  0.01 (V+55)/(1-$\exp$[-(V+55)/10])        &  0.125 $\exp$[-(V+65)/80]  \\
h        &  0.07 $\exp$[-(V+65)/20]                   & 1/($\exp$[-(V+35)/10]+1)
\\ \hline \hline
\end{tabular}
\end{center}
\caption{$\alpha_{X}(V)$ and $\beta_{X}(V)$ functions ($X=m,n,h$)
entering in Eq. (\ref{hh_gating}) for the voltage expressed in
mV.} \label{funz_HH}
\end{table}

We consider the single HH model subjected to $N_E$ (resp. $N_I$) uncorrelated trains
of excitatory (resp. inhibitory) post-synaptic potentials (EPSPs, resp. IPSPs).
Each post-synaptic potential (PSP) is schematized as an instantaneous
variation of the membrane potential by a positive (resp. negative)
amount $\Delta V$ for excitatory (resp. inhibitory) synapses.
Similarly to what has been done in Ref.~\cite{brown_feng_1999},
the amplitude of each voltage kick is assumed to be 0.5 mV, i.e.
reasonably small ($\approx 7$ \%) with respect to the distance
between the "threshold" for spike initiation for rapid EPSPs
and the resting potential ($\approx 6 - 7 $ mV) \cite{noble&stein,tuckwell}.
Moreover, amplitudes $\approx 0.5$ mV are comparable with average
EPSPs experimentally measured for pyramidal neurons in the visual cortex of rats~\cite{koch}.
This amounts to exciting the neuron (\ref{hh},\ref{hh_gating}) with an impulsive current
\begin{equation}
I(t)=Q\Big[\sum_{k=1}^{N_{e}}\sum_{l} \delta(t-t_{k}^{l})
-\sum_{m=1}^{N_{i}}\sum_{n} \delta(t-t_{m}^{n})\Big]
\label{I_imp}
\end{equation}
where $t_{k}^{l}$ (resp. $t_{m}^{n}$) are the arrival times of the
excitatory (resp. inhibitory) PSPs and $Q=C \Delta V$ is the charge
associated to each kick.  The dynamics has been
integrated by employing a fourth order Runge-Kutta scheme with a
time step of $\delta t= 10^{-5} - 10^{-2}$ ms. A spike is
identified when $V(t)$ overcomes a fixed detection threshold
$\Theta = -5$ mV. The results reported in this paper refer to averages performed
over time spans corresponding to $30,000$ to $600,000$ emitted spikes
or (in the low noise limit, where spikes are more rare) at least to an
integration time of $t \approx 1,000 - 10,000 $ s.

In order to reproduce realistic inputs received by cortical
neurons, for each afferent synapse the time interval distribution between PSP
inputs is chosen Poissonian
with an average frequency $\nu_0 = 100$ Hz~\cite{shadlen_newsome_1998}.
Since the trains coming from different neurons are assumed to be
uncorrelated, this amounts to considering only two Poissonian
distributed input trains of kicks of amplitude $\Delta V$, one for
the excitatory and one for the inhibitory neurons with frequencies
$\nu_E = N_e \times \nu_0$ and $\nu_I = N_I \times \nu_0$,
respectively. We consider a number of input neurons of the order of $\sim 100 - 1,000$,
thus the HH neuron is stimulated with average frequencies
$\nu_E (\nu_I) \sim 10^4 - 10^5$ Hz,
consistent with a high-input regime~\cite{shadlen_newsome_1998}.

The stochastic input can be characterized in terms of the net
spike count within a temporal window $\Delta T$
\begin{equation}
N(\Delta T) = \sum_{k=1}^{N_e} n^E_k(\Delta T) - \sum_{m=1}^{N_i} n^I_m(\Delta T)
\label{sp_c}
\end{equation}
where $n^E_k(\Delta T)$ (resp. $n^I_m(\Delta T)$) represents the
number of afferent EPSPs (resp. IPSPs) received from neuron $k$ (resp. $m$)
in the interval $\Delta T$. According to the theory of renewal
processes~\cite{tuckwell} each variable $n(\Delta T)$ has a
gaussian distribution with average $<n(\Delta T)>= \Delta T/
a_{ISI}$ and variance $<n^2(\Delta T)> - <n(\Delta T)>^2 = (\Delta
T \enskip v_{ISI})/a^3_{ISI}$, where $a_{ISI}$ and $v_{ISI}$
indicate average and variance of the ISI-distribution. In particular
for Poissonian distributed ISIs with average frequency $\nu_0$:
$a_{ISI}=1/\nu_0$ and $v_{ISI}=1/\nu_0^2$.

By assuming statistically independent input trains, also $N(\Delta(T))$
follows a gaussian distribution with average and variance given by
\bey
<N(\Delta T)> &=& (N_e - N_i) \frac{\Delta T}{a_{ISI}}
\quad,
\label{ave_var}
\\
Var[N(\Delta T)]&=&(N_e+N_i)
\frac{(\Delta T \enskip v_{ISI})}{a^3_{ISI}}=\sigma^2 \frac{\Delta T}{a_{ISI}}
\enskip .
\nonumber
\eey
Here it has also been assumed  that the excitatory and inhibitory inputs are
characterized by the same $a_{ISI}$ and $v_{ISI}$. The parameter
$\sigma=(\sqrt{(N_e+N_i) v_{ISI}})/a_{ISI}$ measures the standard deviation
of the stochastic input process.

 In the high-input regime the distribution of $N(\Delta(T))$
turns out to be the same for input ISIs following a Poissonian
or a uniform distribution
\begin{equation}
P_{U}(t)=\left\{ \begin{array}{ll}
\frac{\nu_0}{2\varepsilon} & \mbox{if $\frac{(1-\varepsilon)}{\nu_0}\le t \le \frac{(1+\varepsilon)}{\nu_0}$}     \\
0 & \mbox{otherwise}
\end{array} \right.
\end{equation}
provided that both distributions have the same average and variance. For
$P_{U}(t)$, with $0\le \varepsilon \le 1$, $a_{ISI} = 1/\nu_0$,
$v_{ISI}=\varepsilon^2/(\nu_0^2 3)$ and $\sigma = \varepsilon \sqrt{(N_e + N_i)/3}$.
Since we are interested in analyzing the effect of the variance of
the stochastic process on the response of the HH model
for fixed $<N(\Delta T)>$, we will consider uniformly distributed
ISIs, because this kind of distribution allows to reach extremely small
values of the variance (by lowering the $\varepsilon$-parameter) without
modifying the average.

Both for Poissonian and uniform distributions with fixed $a_{ISI} = 1/\nu_0$
the average current stimulating the neuron is given by
\begin{equation}
\bar I= \frac{C \Delta V <N(\Delta T)>}{\Delta T} = C \Delta V
\nu_0 (N_e - N_i).
\label{I_ave}
\end{equation}
The bifurcation parameter $\bar I$ determines in which dynamical
regime the neuron is operating.

\begin{figure}[h]
\begin{center}
\includegraphics*[clip,width=7cm]{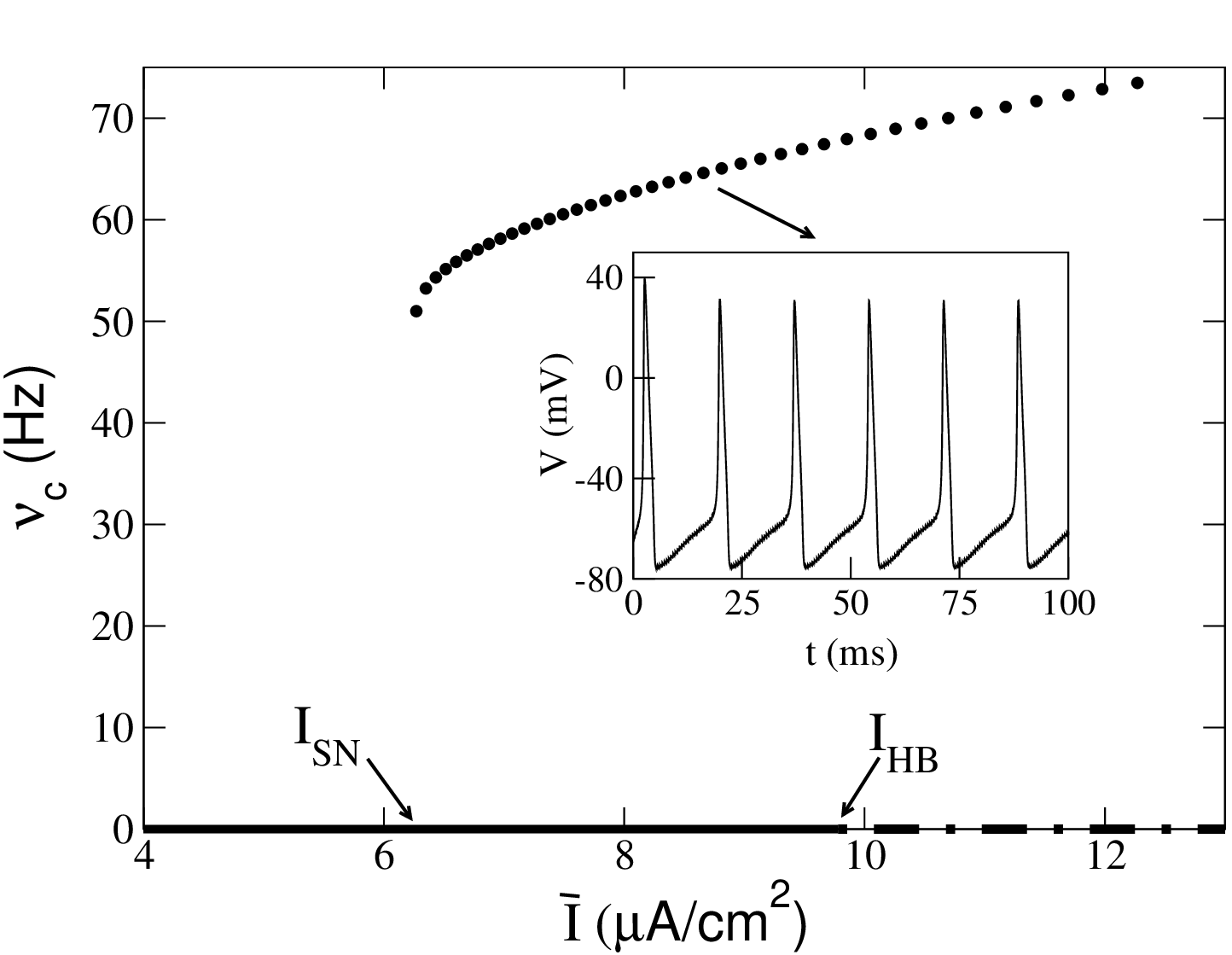}
\end{center}
\caption{The filled circles represent the
frequency-current response curve ($\nu_C$ versus $\bar I$)
obtained by considering a periodic train of kicks of amplitude
$\Delta V=0.5$ mV and frequency $\bar \nu = \bar I /(C \Delta V)=\nu_0 (N_e-N_i)$.
The solid (resp. dash-dotted)
line indicates the stable (resp. unstable) fixed point solution.
In the inset the time evolution of the membrane potential $V(t)$,
corresponding to a typical limit cycle solution, is shown.
}
\label{F_I}
\end{figure}

Since it is equivalent to stimulate a neuron with a constant current $I(t) \equiv I_{dc}$ or
with a periodic train of kicks of sufficiently high frequency $\bar \nu$,
the frequency-current response curve (usually obtained with a constant
input current) can also be recovered by considering a periodic input originating
from $N_i$ inhibitory and $N_e$ excitatory synapses each firing with a frequency $\nu_0$
(cf. Fig. \ref{F_I}).

Three principal dynamical regimes can be singled out for the HH
model subject to a constant current stimulation
\cite{rinzel_miller_1980}. For small currents the HH model is in a silent regime, i.e. its
dynamics is always attracted by a stable fixed point.
In particular, the relaxation towards this state is characterized by
damped oscillations, since it is a focus. By increasing the current
the fixed point looses its stability via a sub-critical Hopf bifurcation
for $\bar I = I_{HB} \simeq 9.78 \mu$A/cm$^2$. The unstable limit cycles emerging at $I_{HB}$
annihilates via a saddle-node bifurcation with stable periodic
oscillations, corresponding to tonic firing, at a lower value of the
current $\bar I = I_{SN} \simeq 6.27 \mu$A/cm$^2$.
Therefore the following three
dynamical regimes can be identified:

\begin{enumerate}

\item a silent regime for $
\bar I < I_{SN}$;

\item a bistable regime, where the fixed point
coexists with a stable limit cycle solution for $ I_{SN} < \bar I <
I_{HB}$;

\item a periodic firing regime for $ \bar I >
I_{HB}$.

\end{enumerate}

\subsection{Statistical and dynamical indicators}

In order to characterize the output of the neuron and
to examine the coherence effects in the response we
have employed the following indicators:
\begin{itemize}
\item the distribution of the output interspike intervals
$P_{ISI}(t)$ and its first moments: the average ISI ($A_{ISI}$)
and the corresponding standard deviation ($S_{ISI}$);

\item the spike-triggered average potential (STAP)~\cite{bialek}
that gives the average shape of the membrane potential preceding
the emission of a spike and the spike-triggered average input
fluctuations (STAF): \be q(t)=\frac{(N_t (\Delta T) -<N(\Delta
T)>)}{\sqrt{Var[N(\Delta T)]}} \label{corr} \ee where $N_t (\Delta
T)$ is the net spike count at time $t < 0$ before a spike emission
occurring at $t=0$, while the expressions of $<N(\Delta T)>$ and
$Var[N(\Delta T)]$ are reported in Eqs. (\ref{ave_var}). The STAF
(\ref{corr}) can be related to current fluctuations, estimated
over a time window $\Delta T$, with respect to their average value
$\bar I$ (see Eq. (\ref{I_ave})). A positive (resp. negative)
value of $q$ at a certain time $t < 0$ indicates a correlation
between a positive (resp. negative) current fluctuation at that
time and the emission of a spike at $t=0$.

\item the coefficient of variation of the ISIs
\begin{equation}
R=\frac{S_{ISI}}{A_{ISI}}
\quad ,
\label{CV}
\end{equation}
typically employed to characterize the nature of a process, being
$R=0$ for a perfectly periodic response and $R=1$ for Poissonian output;

\item the normalized autocorrelation function
$C(t)$ for the membrane potential and the
correlation time~\cite{pikovsky_kurths_1997} defined as
\begin{equation}
\tau_c= \int_0^\infty C^2(t) \enskip dt
\quad .
\label{tauc}
\end{equation}

\end{itemize}

As a further indicator we have
employed the {\it conditional entropies} $h(N)$ \cite{shannon}.
In order to define these quantities, we need to digitize the
output of the neuron in a binary sequence, where $1$ and $0$ indicates
respectively presence or absence of a spike in a certain time window $\Delta t$.
The choice of the resolution $\Delta t$ employed to
analyze the output is crucial \cite{cover}~\footnote{The time window
$\Delta t$ should be chosen in an appropriate
way to avoid on one side to observe two spikes within the same
window and on the other side to have extremely long series of
zeros, that would spoil the statistical analysis.
A good choice of $\Delta t$ is the refractory period of the
neuron following the emission of a spike and an upper bound for
this time can be given by the measured minimal ISI value. Since
the period of relative refractoriness decreases upon increasing
the amount of noise, for each current we have chosen $\Delta t$ to be
the minimal ISI measured for the maximum noise variance considered.
As a matter of fact $\Delta t=5$ ms has been used for all current values,
apart from ${\bar I}=0$ where $\Delta t=7$ ms was more appropriate.}.

Indicating with $s_k$ the binary symbol associated to the $k$-th
window of the time series and with $W_N=(s_{i+1},s_{i+2},\dots,s_{i+N})$ a
sequence of symbols (word) of length $N$, the probability that a
certain word will be observed is $P(W_N)$. The Shannon block
entropy is then defined as
\begin{equation}
H(N)=-\sum_{\{W_{N}\}} P(W_{N})\log_2 P(W_{N})
\label{block_entropy}
\end{equation}
where the sum is extended to all the $2^N$ possible words of
length $N$. The conditional entropies are given by the following
difference~\cite{cover}
\bey
&&h(N)=H(N+1) - H(N)=
\label{cond_ent}
\\
&=&-\left< \sum_{\{s_{N+1}\}} P(s_{N+1}|W_{N})\log_2 P(s_{N+1}|W_{N}) \right>_{\{W_{N}\}}
\nonumber
\eey
where the brackets indicate the average over all possible
sequences of length $N$ preceding the symbol $s_{N+1}$ and
$P(s_{N+1}|W_{N})$ is the conditional probability to observe
symbol $s_{N+1}$ once the word $W_{N}$ has been registered.
By definition we set $h(0)=H(1)$.

The conditional entropies represent the average information gained by
the knowledge of the $(N+1)$-th symbol of a time series, once the
other $N$ symbols are already known.
For sufficiently long words, if the examined
process can be considered as ``ergodic'', $h(N)$ tends to an
asymptotic value $h_A$, the length $M$ for which the saturation is
attained gives the order (the memory) of the Markovian process
able to reproduce the considered dynamics~\cite{farmer}.
For a regular process (e.g., a periodic state) $h_A=0$, while for a
purely stochastic process the conditional entropies attain
their maximal value, i.e. $h_A=1$ bit\footnote{From a numerical
point of view it is known that
the estimation of Shannon entropy (\ref{block_entropy}) from
finite samples can lead to underestimation of $H(N)$. In order
to reduce the systematic errors arising from naive estimations
of $P(W_N)$ we have employed the analytic estimator recently
introduced by Grassberger \cite{grass_estimator}.}.

As already mentioned,
the conditional entropies have been successfully employed to characterize
SR~\cite{neiman_1996}. In particular a clear minimum for $h_A$ was observed
in correspondence to an optimal noise amplitude
giving rise to a maximum in the signal-to-noise ratio (SNR), that
is a typical signature of SR. The minimum in $h_A$
was associated with the most
ordered (coherent) structure of the output binary sequence.

Finally, in order to characterize the correlations present
in the binary sequence
we use the following autocorrelation function ~\cite{koch}
\begin{equation}
C_{bin}(k)= \frac{\sum_i \left( <s_i s_{i+k}> -<s_i> \right)}{\sum_i \left( <s_i^2> -<s_i>^2 \right)}
\label{corr_bin}
\end{equation}
and the associated correlation time $\tau_{bin}$ defined as
in Eq. (\ref{tauc}).

In the next sections we investigate the response of the HH model
stimulated by stochastic spike trains in the three dynamical
regimes shown in Fig.~\ref{F_I}.

\section{Silent regime}
\label{tre}

\begin{figure}[h]
\begin{center}
\includegraphics*[clip,width=7cm]{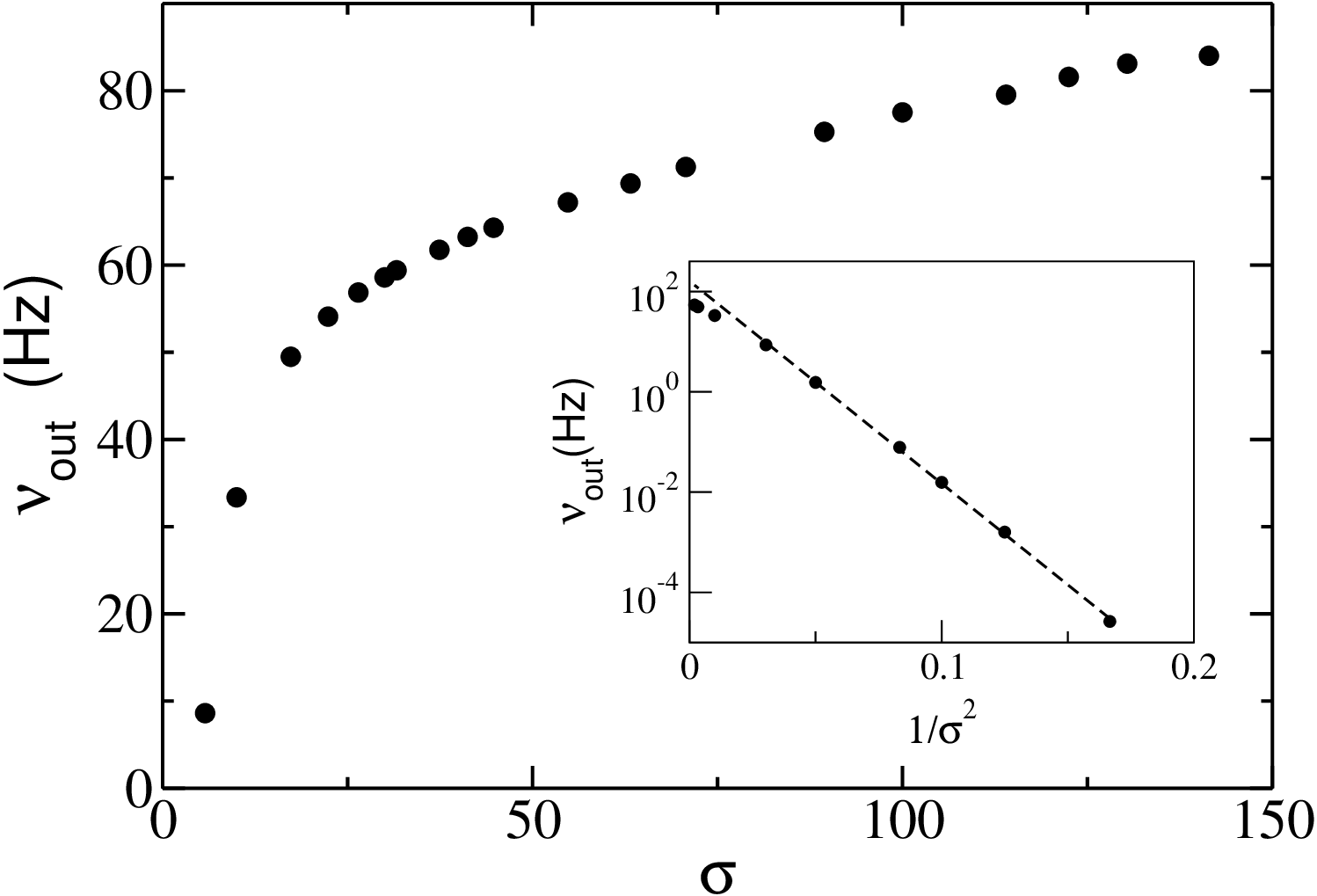}
\end{center}
\caption{Average firing rate $\nu_{out}$ as a function of $\sigma$
for $\bar I = 5 \mu$A/cm$^2$. In the inset the frequency is
plotted in a lin-log scale as a function of $1/\sigma^2$ for
$\sigma^2 < 500$. The dashed line is an exponential fit
$y=A*\exp{[-\sigma^2_c/\sigma^2]}$ in the range $6 <  \sigma^2 < 20 $
with $A=162$ and $\sigma^2_c \approx 93$.} \label{i5_fout}
\end{figure}

In the regime where the neuron subjected to a constant current
does not fire, i.e. in the range $\bar I < I_{SN} \simeq
6.27 \mu$A/cm$^2$, the presence of noise in the input
(characterized by the standard deviation of the noise $\sigma$)
induces stochastic firing of the model with an average firing
rate $\nu_{out} \equiv 1/A_{ISI}$ steadily increasing with $\sigma$ (as shown in Fig. \ref{i5_fout}
for $\bar I = 5 \mu$A/cm$^2$).

\begin{figure}[h]
\begin{center}
\includegraphics*[clip,width=5cm]{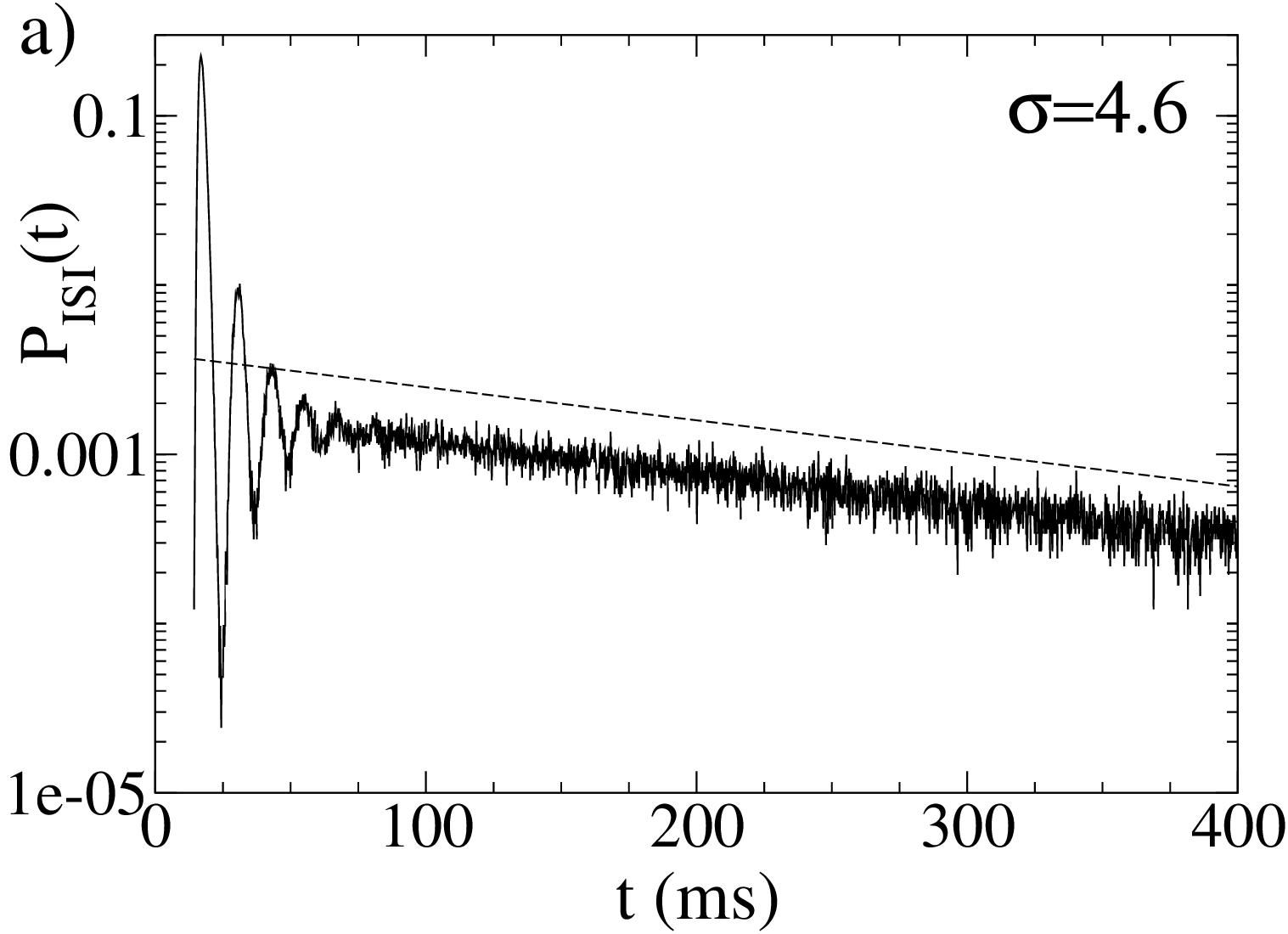}
\includegraphics*[clip,width=5cm]{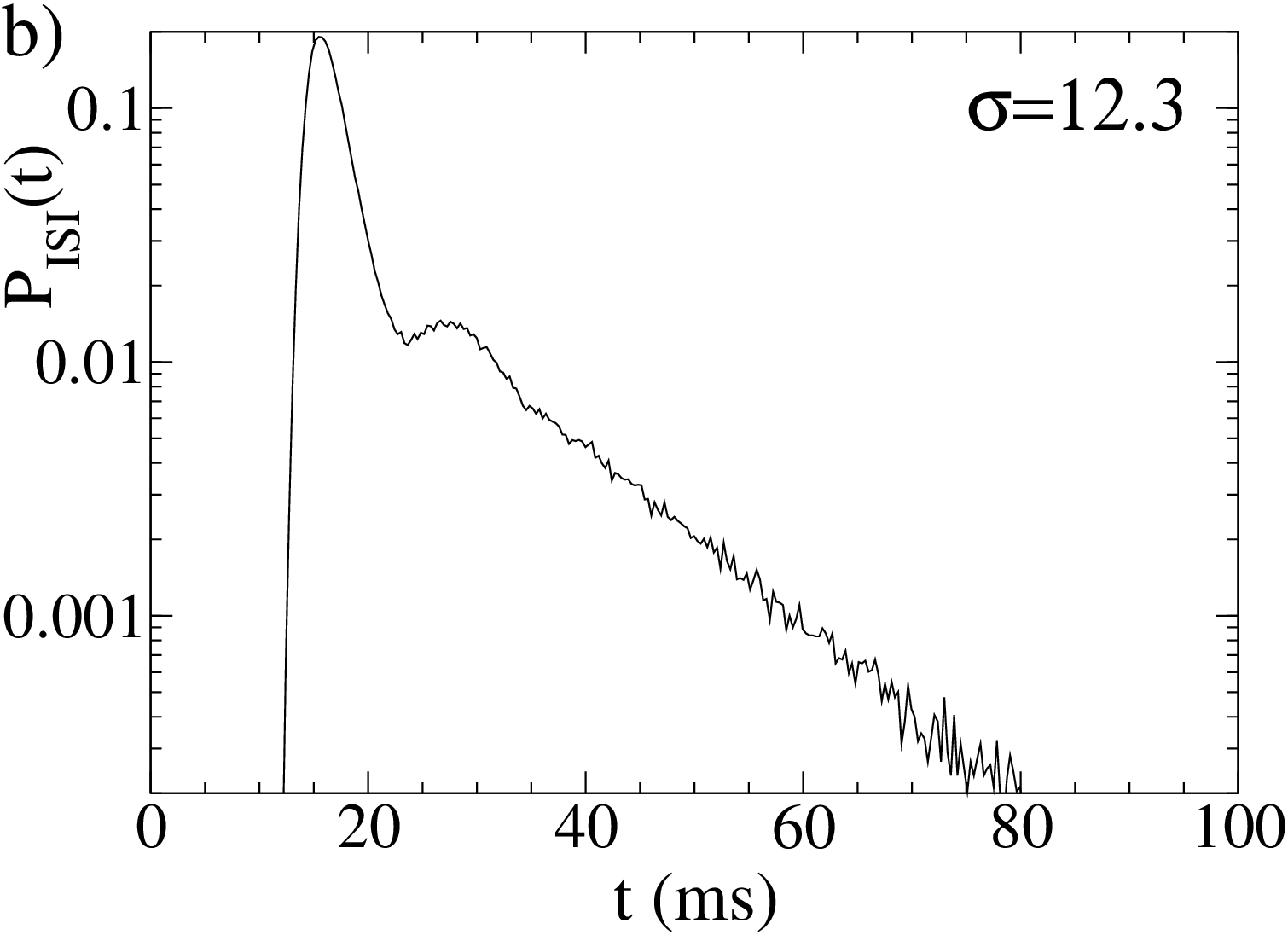}
\includegraphics*[clip,width=5cm]{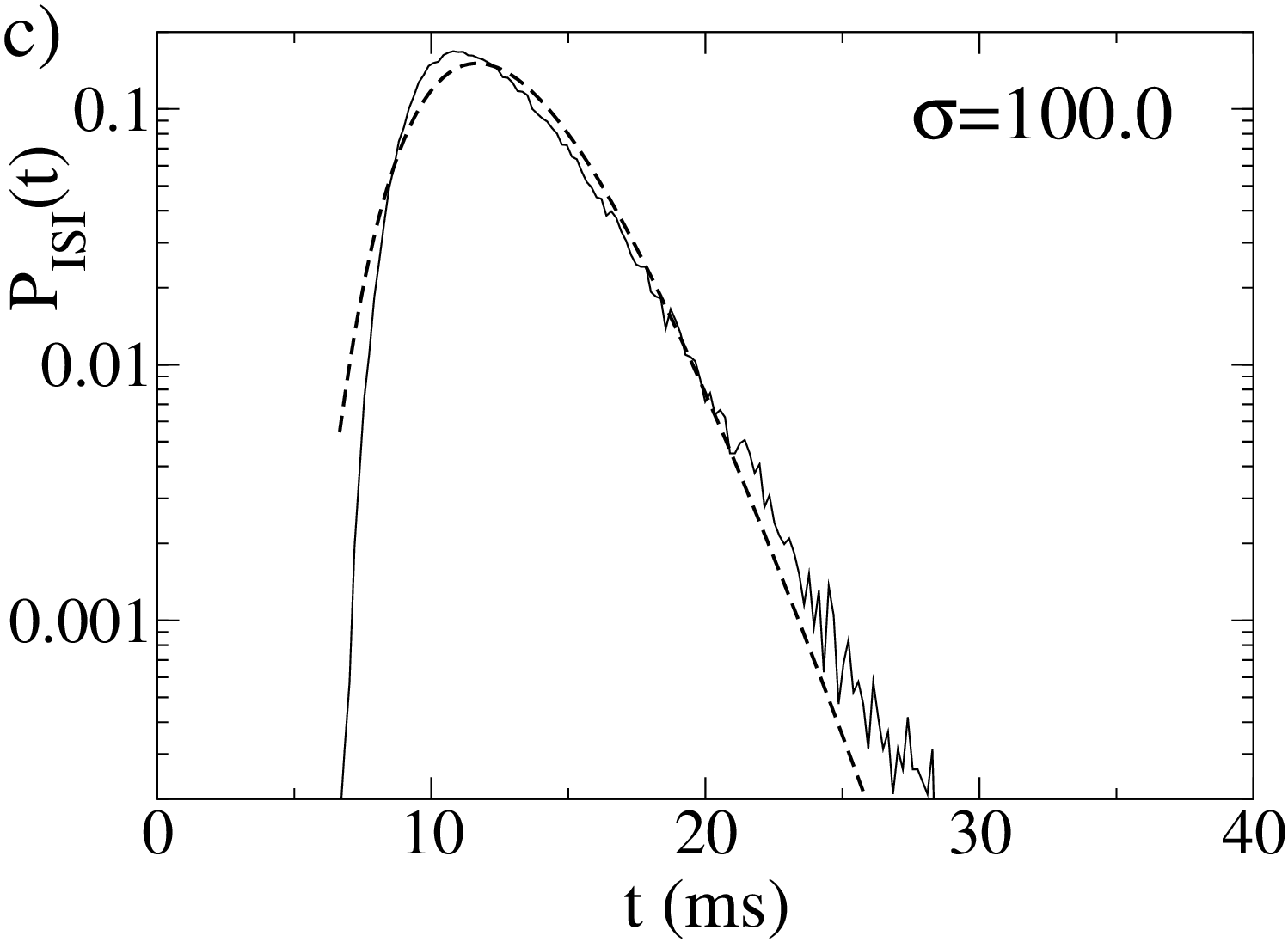}
\end{center}
\caption{Probability density distribution $P_{ISI}(t)$ for $\bar I
= 6.15 \mu$A/cm$^2$. The figures refer to three different values
of $\sigma$: (a) 4.6, (b) 12.3, and (c) 100.0. The dashed line in
(a) indicates the slope of the exponential tail, while in (c) it
represents an inverse Gaussian distribution with the same average
and the same variance as the original $P_{ISI}(t)$. }
\label{pisi_i6.15}
\end{figure}

For a fixed average input current $\bar I$ the ISI-distribution
depends strongly on the standard deviation $\sigma$ of the noise (cf. Fig.
\ref{pisi_i6.15}). For low noise $P_{ISI}(t)$ exhibits a
multimodal structure with an exponential tail, for increasing
noise the additional peaks and the tail disappear gradually. In
the high noise limit $P_{ISI}(t)$ tends to an inverse Gaussian~\cite{invgau,tuckwell}.

 In the following sub-section \ref{tre1}
 we discuss the mechanisms responsible for the firing activity
 of the neuron in the low and high noise limits, while
 sub-section \ref{tre2} is devoted to coherence resonance
 phenomena observed for intermediate level of noise fluctuations.

\subsection{Response of the neuron to low and high noise}
\label{tre1}

Let us discuss the origin of the main features
present in $P_{ISI}(t)$ at low noise. The coexistence of
multi-peaks and the exponential tail in  Fig.~\ref{pisi_i6.15}a
suggests that at least two different mechanisms are responsible
for the firing of the neuron.

\subsubsection{Spikes triggered by relaxation oscillations}

We can safely affirm that the multiple peaks in the $P_{ISI}(t)$ are associated
with the relaxation oscillations of the membrane potential towards the rest
state following the emission of a spike. However, a linear stability analysis around
the stable fixed point can provide the oscillation periods around the focus,
but is not sufficient to fully characterize the relaxation dynamics, that exhibits also
nonlinear aspects.

\begin{figure}[h]
\begin{center}
\includegraphics*[width=6.9cm]{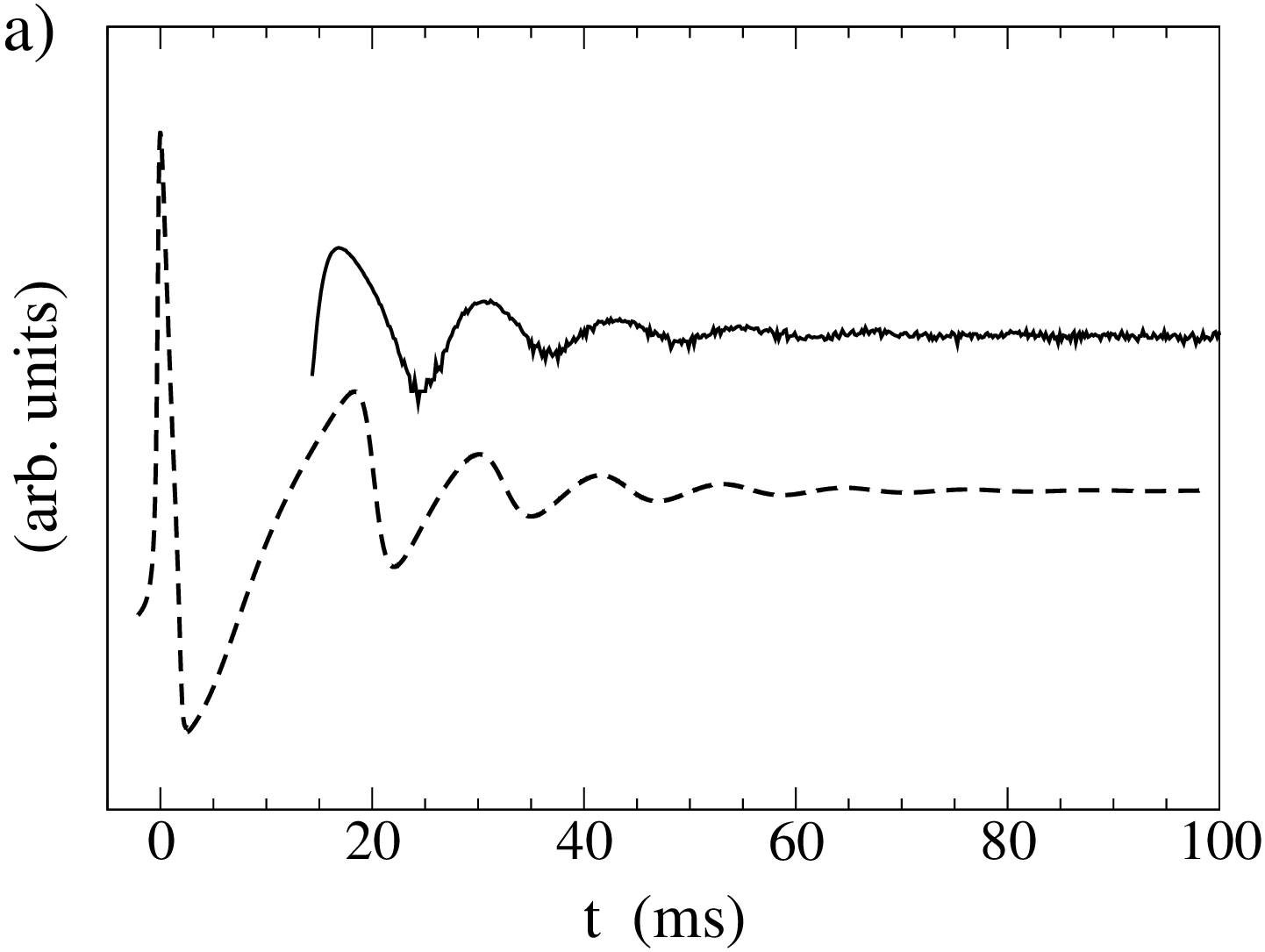}
\includegraphics*[width=7cm]{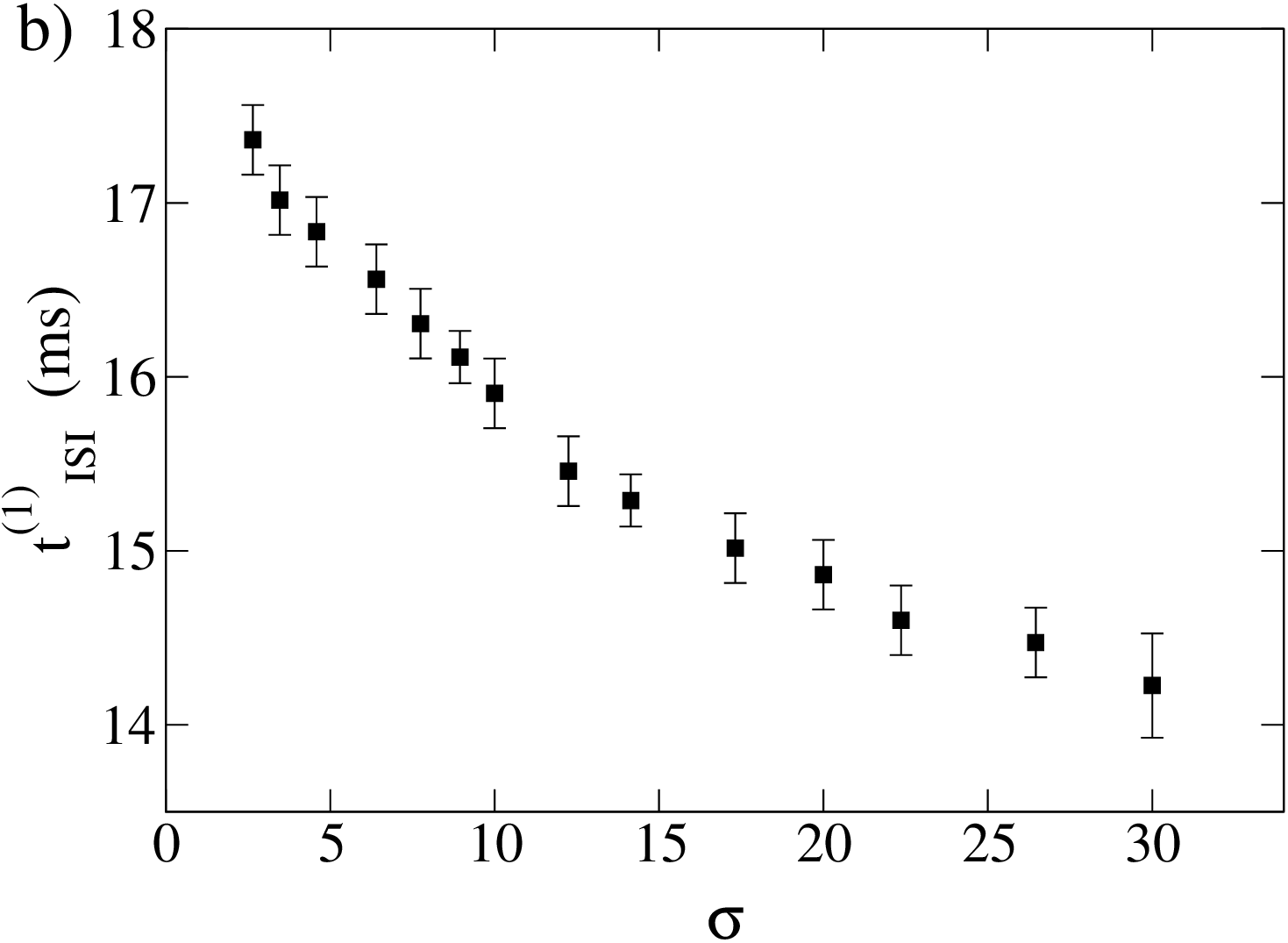}
\end{center}
\caption{(a) $P_{ISI}(t)$ obtained by the stochastic input
(\ref{I_imp}) with $\sigma=4.6$ (continuous curve) and the
potential output $V(t)$ following a step current stimulation
(\ref{step}) (dashed line). The
position of the spike has been shifted to $t=0$ and
the action potential has been rescaled to better
reveal the relaxation oscillations. (b) $t^{(1)}_{ISI}$
as a function of $\sigma$. The error
bars correspond to the histogram resolution employed to estimate
$P_{ISI}(t)$. Data refer to $\bar I =  6.15 \mu$A/cm$^2$.
} \label{multi}
\end{figure}

In order to understand if these oscillations are responsible
for the multimodal structure of $P_{ISI}(t)$,
we have stimulated the model (\ref{hh},\ref{hh_gating}) with a
step current of amplitude $\bar I$, i.e.
\begin{equation}
I(t)=\left\{
\begin{array}{ll} 0  &
\mbox{if  $t \le 0$}     \\
\bar I
& \mbox{if $t>0$}
\end{array} \right.
\qquad ,
\label{step}
\end{equation}
and registered the shape of the output $V(t)$. As shown in Fig.
\ref{multi}a (dashed line), the HH model responds emitting
one or more action potentials followed by damped oscillations.
The first oscillation of period $T_{nl}$ has a clear nonlinear character,
while we have verified that the angular frequency $2 \pi/T_l$ describing the
subsequent oscillations corresponds to the imaginary part of the complex conjugate
Floquet eigenvalues associated to the stable fixed point.
Therefore the latter solutions can be completely characterized
within a linear stability analysis of model (\ref{hh},\ref{hh_gating}).

\begin{figure}[h]
\begin{center}
\includegraphics*[clip,width=7cm]{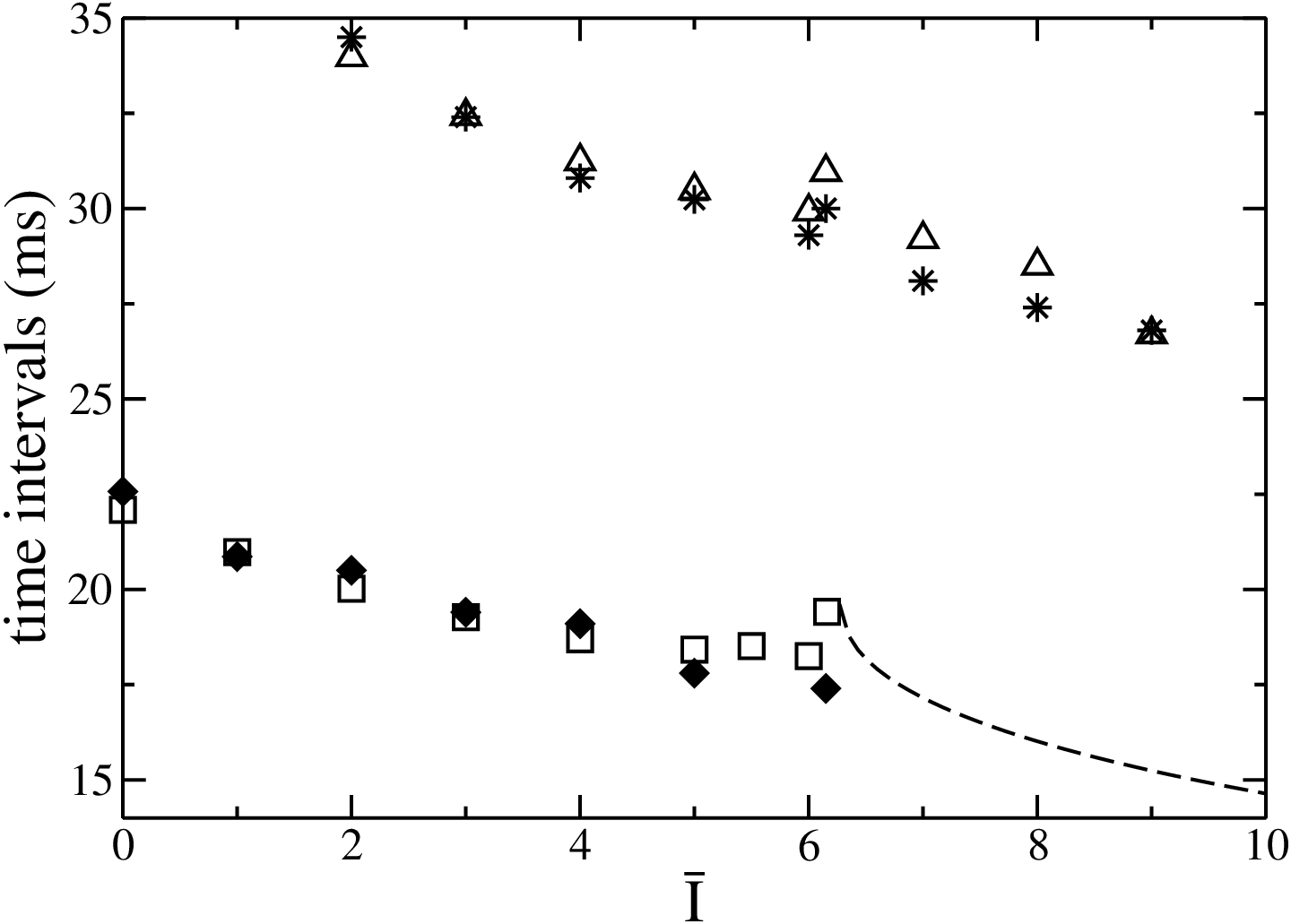}
\end{center}
\caption{Positions of the peaks of the distribution $P_{ISI}(t)$ as
a function of the average current $\bar I$ for low noise:
first peak $t^{(1)}_{ISI}$ (filled diamonds) and second peak
$t^{(2)}_{ISI}$ (asterisks). For comparison the
time intervals separating the induced spike from the maximum associated to
the first (nonlinear) oscillation $\tilde T_{nl}$ (empty squares) and to the
second (linear) oscillation $\tilde T_{nl} + T_l$ (empty triangles) are shown.
The periods $1/\nu_c$ of the regular spike trains emerging
above $I_{SN}$ are also plotted (dashed line).
The data for $t^{(1)}_{ISI}$ and $t^{(2)}_{ISI}$ have been obtained for
$\sigma \approx 2.6 - 10$.
}
\label{peaks}
\end{figure}

To better compare the periods of the relaxation oscillations
with the measured positions of the first (resp.
second) peak $t^{(1)}_{ISI}$ (resp. $t^{(2)}_{ISI}$) of the ISI distribution,
we have registered the temporal interval separating
the peak of the action potential and the first subsequent maximum $T_{nl}$
and corrected (increased) these values by the corresponding rise time needed
to reach the detection threshold $\Theta$. As shown in Fig.~\ref{peaks},
the comparison of $t^{(1)}_{ISI}$ with the corrected values $\tilde T_{nl}$
is very good for $0 \le \bar I \le 5 \mu A/cm^2$. Moreover, by approaching
$I_{SN}$ we observe that $\tilde T_{nl} \to 1/\nu_c$ (where
$\nu_c$ is the frequency of the spike train limit cycles
emerging at the saddle-node transition).
For completeness we have also
verified that the other peaks observed in the distribution $P_{ISI}(t)$
are related to the subsequent linear oscillations of period $T_l$
shown in Fig. \ref{multi}a (dashed line). In particular, we
concentrated on the second peak $t^{(2)}_{ISI}$ and compared it with
$\tilde T_{nl} + T_l$ , also this time the agreement is quite good
(see Fig.~\ref{peaks})\footnote{Above the saddle-node transition
$\bar I > I_{SN}$ the position of the second peak has been
estimated as $1/\nu_c + T_l$}. In general
for the $m$-th peak we expect that a good approximation of its
position will be given by $\tilde T_{nl} + (m-1) T_l$, and
this is indeed verified as shown in Fig. \ref{multi}a.

The damped oscillations following a spike induce a modulation
in the degree of excitability of the model, or analogously
a modulated effective threshold for spike elicitation. Indeed
a $P_{ISI}(t)$ quite similar to that displayed in Fig. \ref{multi}a
was obtained by Wilbur \& Rinzel by considering a leaky
integrate-and-fire model with a threshold evolving
dynamically in time~\cite{wilbur_rinzel_1983}.

So far we limited ourselves to low
values of noise variance, by increasing $\sigma$ we expect that the
position of the first peak will be shifted to smaller values.
Due to stronger noise fluctuations,
after the emission of a spike a second spike can be elicited even
before that the subsequent relaxation oscillation of $V(t)$ reaches its
maximum value. This tendency of $t^{(1)}_{ISI}$ to
decrease with $\sigma$ is illustrated in Fig. \ref{multi}b.

\subsubsection{Firing activated by noise}

In the first part of this subsection
our analysis has been devoted to relatively short ISIs,
however the exponential tail
present in $P_{ISI}(t)$ (see Fig.~\ref{pisi_i6.15}a)
is extremely relevant since it gives the main contribution
to the average firing rate $\nu_{out}$ of the neuron. As already
mentioned, this long lasting tail should be related to some sort of
activation process.

\begin{figure}[h]
\begin{center}
\includegraphics*[clip,width=7cm]{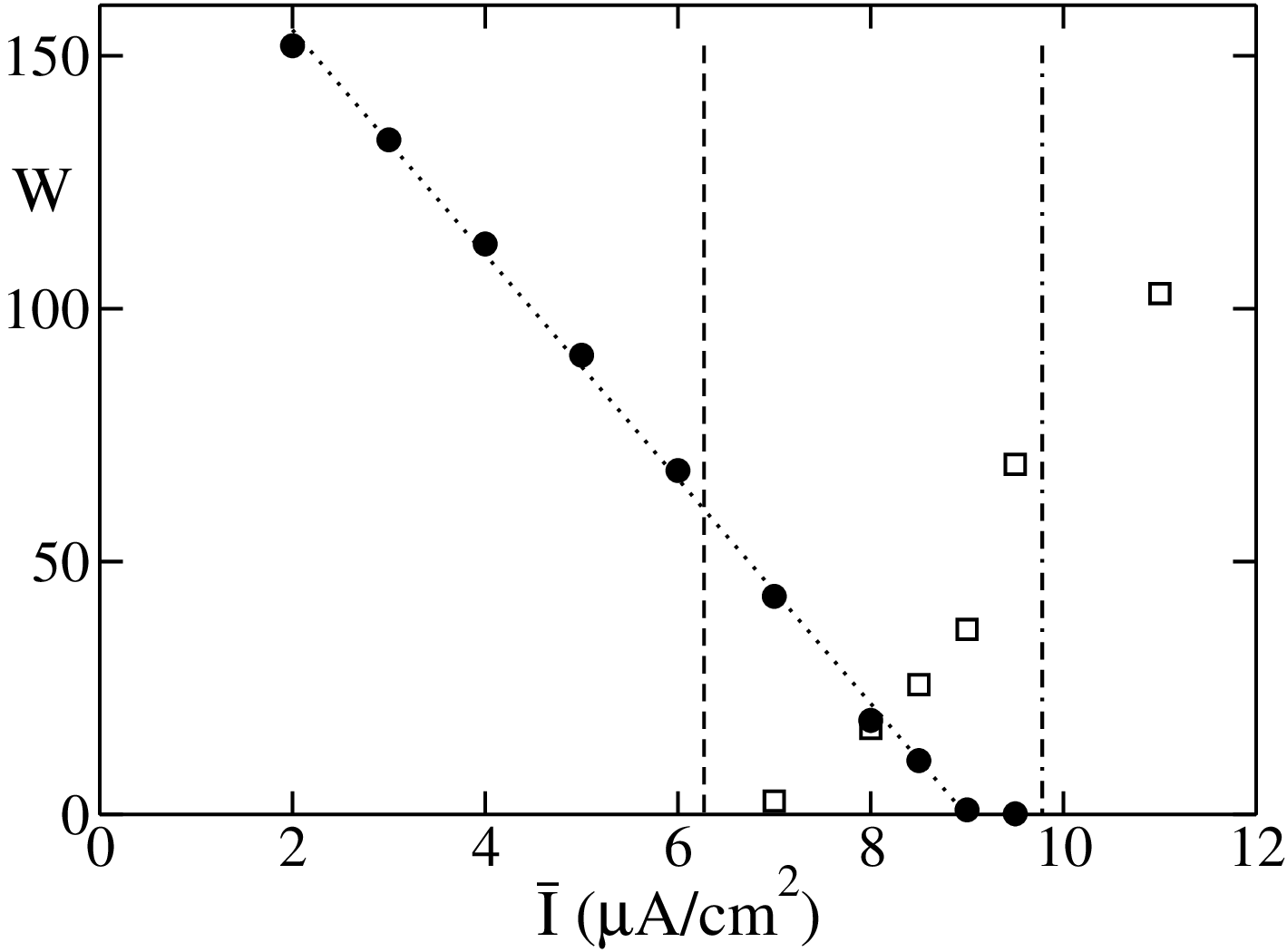}
\end{center}
\caption{Activation barrier heights $W$ as a function of
the average input current for the silent state $W_S$ (filled circles)
and the oscillatory state $W_O$ (empty squares). The vertical dashed
(resp. dot-dashed) line indicates the position of
$I_{SN}$ (resp. $I_{HB}$). A linear fitting to  $W_S$
in the interval $ 2 \mu A/cm^2 \le \bar I \le 9 \mu A/cm^2$ is also
shown as a dotted line.
} \label{barrier}
\end{figure}

An activated firing can arise in this regime due to the
competition between the tendency of the HH dynamics to relax
towards its stable fixed point ($V_{rest}(\bar I)$)
and noise fluctuations that instead lead
the system towards an excitation threshold (note, however, that there is no
fixed threshold in the HH-model~\cite{tuckwell}). Therefore, the
dynamics of $V(t)$ resembles the overdamped dynamics
of a particle in a potential well under the influence of thermal
(stochastic) fluctuations~\cite{kramers}. Due to the activation
process the membrane potential can be driven towards the excitation
threshold with an average escape time
given by Kramers expression~\cite{kramers}
\begin{equation}
T_e \propto {\rm e}^{W_S/\sigma^2}
\quad ,
\label{rate}
\end{equation}
where $W_S$ plays the role of an energy barrier and $\sigma^2$ of
an effective temperature of the bath. This behavior is indeed
verified for small $\sigma$-values as shown in the inset of Fig.
\ref{i5_fout}. Therefore we expect that for $\sigma < \sigma_c
\equiv \sqrt{W_S}$ ($\sigma_c \approx 9.64$ for $\bar I=5
\mu$A/cm$^2$) the dynamics can be characterized as an activation
process, while for $\sigma^2$ larger than the barrier value the
dynamics should be mainly diffusive. A further indication that for
low noise the dynamics is essentially Poissonian is given by the
fact that $R \approx 1$ for $\sigma < \sigma_c$. We have estimated
$W_S$ for various values of the average current for $\bar I <
I_{HB}$ and a linear decrease of the barrier height with $\bar I$
is clearly observable in Fig. \ref{barrier}, except in direct
proximity of $I_{HB}$. Moreover (as expected) $W_S \to 0$
approaching $I_{HB}$ where the fixed point looses its stability
via a sub-critical Hopf bifurcation \footnote{The average escape
times $T_e$ have been estimated directly as $1/\nu_{out}$ for $
\bar I < I_{SN}$ since at sufficiently low noise the activation is
the prevailing mechanism in this regime; above $I_{SN}$ where
there is a coexistence of a stable limit cycle with a stable fixed
point these times have been measured as the residence times $T_S$
in the silent state (i.e. in proximity of the fixed point).}.

\begin{figure}[h]
\begin{center}
\includegraphics*[clip,width=7cm]{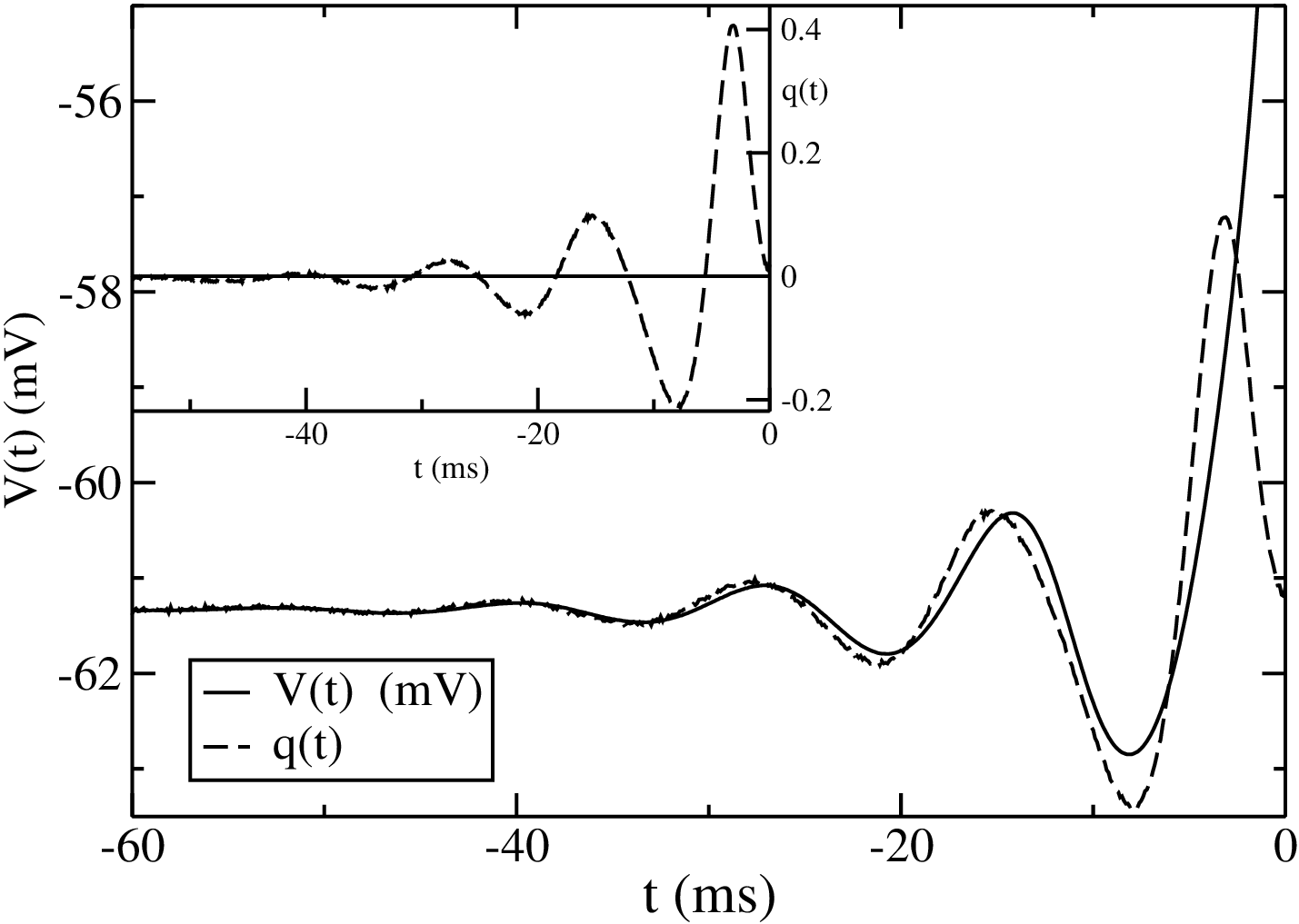}
\end{center}
\caption{STAF $q(t)$ (dashed line) and
STAP $V(t)$ (solid line)
preceding the emission of a spike.  At time $t=0$
the potential overcomes the detection threshold $\Theta$.
Both quantities have been defined by employing a time window
$\Delta T = 0.1$ ms and have been averaged only over ISIs longer
than 150 ms.
The data refer to $\bar I= 5 \mu A/cm^2$ and $\sigma=5.7$.
For a better comparison $q(t)$ has been shifted and amplified by
a factor 10, the original function is displayed in the inset.}
\label{pot}
\end{figure}

To gain some deeper insight into the role of noise fluctuations
in eliciting a spike, we have estimated
the STAP $V(t)$ and the STAF $q(t)$ for sufficiently long ISIs and for
small $\sigma$.
As shown in Fig.~\ref{pot},
$q(t)$ and $V(t)$ exhibit oscillations of period $\approx T_l$
preceding the spike emission and these oscillations are
almost in phase, apart from a small delay
in the response of $V(t)$ with respect to the input noise.
This indicates that the potential follows the current oscillations
and that the emission of a spike (for long ISIs) is commonly triggered by
the excitation of linear subthreshold oscillations around the rest potential.
Therefore the neuron in proximity of the rest state acts as a sort of selective filter
since it responds (by emitting a spike) with higher probability when excited with a specific input
frequency ($\approx 1/T_l=61-88$ Hz for $ 0 \le \bar I \le I_{SN}$).
This result agrees with a previous analysis reported in~\cite{yu_wang_2001},
where it has been shown that a silent HH neuron, subjected
to a sinusoidal current, optimally resonates when forced with a
frequency linearly correlated with $1/T_l$.

\subsubsection{High noise limit}

As already mentioned, for sufficiently high
noise the probability density distribution of the ISIs
reduces to an inverse Gaussian. The
reason is that for $\sigma \gg \sigma_c$ the ``potential well''
surrounding the stable fixed point becomes irrelevant for the
dynamics due to the amplitude of the noise fluctuations,
whose effect on the dynamics of the neuron are twofold:
a constant current $\bar I$ driving the
system plus a stochastic (Gaussian) term with zero average. Therefore we can assume
that the dynamics of the membrane potential is determined by a Langevin
equation with a drift term. The distribution of the first passage
times from a certain threshold, in presence of a reset mechanism
to a fixed rest potential, is given by the well known inverse
Gaussian distribution~\cite{invgau,tuckwell}
\begin{equation}
f(t)=\frac{\alpha}{\sqrt{2 \pi \beta t^3}}
{\rm e}^{-\frac{(t-\alpha)^2}{2 \beta t}}
\quad ,
\label{inv_gau}
\end{equation}
where $\alpha = A_{ISI}$ and $\beta=S_{ISI}^2/A_{ISI}$. The
good agreement shown in Fig. \ref{pisi_i6.15}c confirms that the
mechanism leading to repetitive firing in the high noise limit can be
indeed schematized as a Wiener process plus drift.

\begin{figure}[h]
\begin{center}
\includegraphics*[clip,width=7cm]{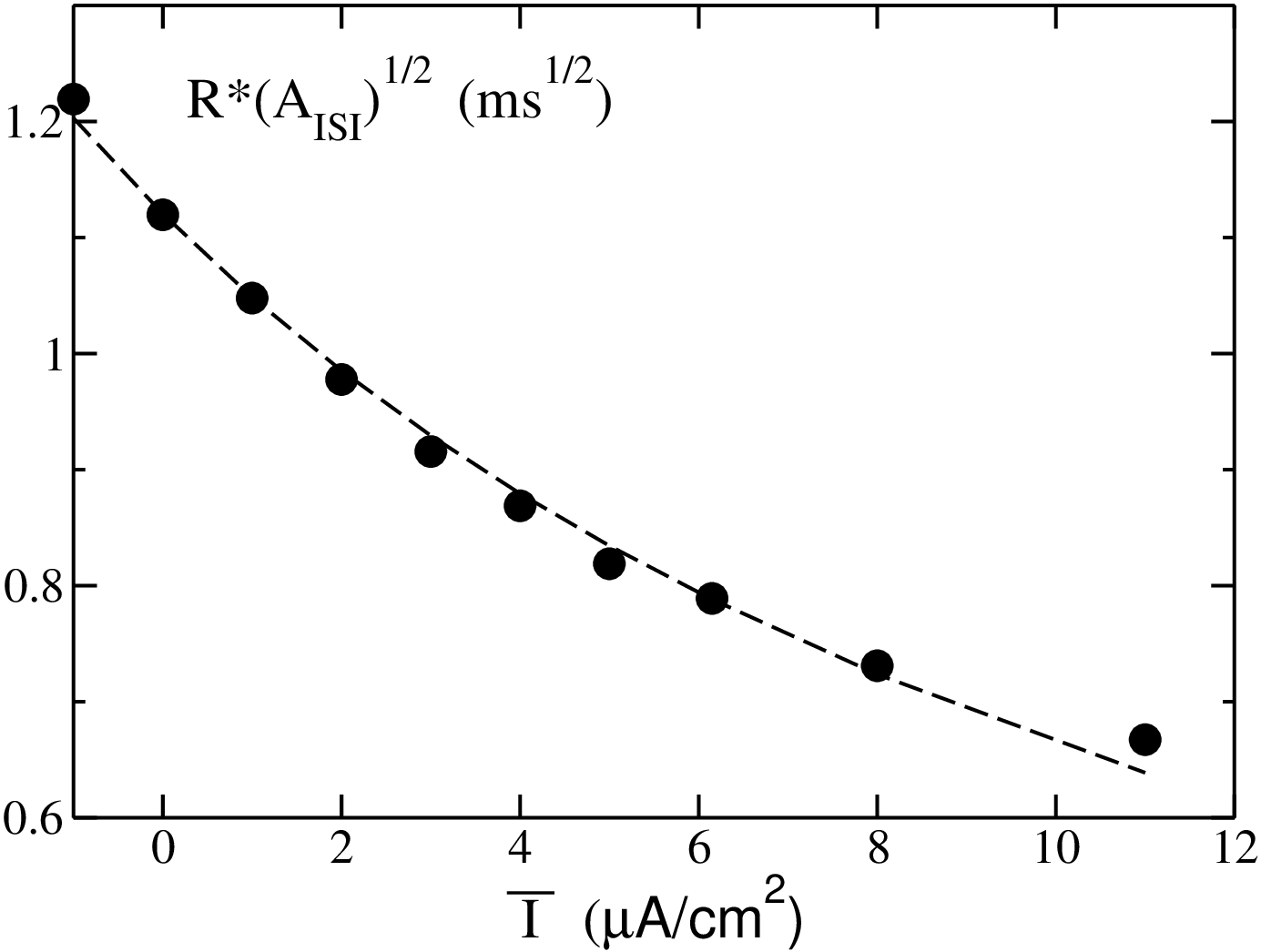}
\end{center}
\caption{$R \times \sqrt{A_{ISI}} $ (filled circles)
versus $\bar I$, the dashed line is a best fit to the data
of the form $a/(\bar I + I_0)$ with $a=16.35 \enskip ms^{1/2}\mu A/cm^2$
and $I_0=14.59 \enskip \mu A/cm^2$.
The data refer to $\sigma=100$.
}
\label{cvisi}
\end{figure}

Moreover, by assuming that the drift is proportional to $\bar I$ and the
amplitude of the effective noise to $\sigma$ one obtains the
following dependence for the ISI coefficient of variation~\cite{tuckwell}
\begin{equation}
R \propto \frac{\sigma}{(\bar I + I_0)\sqrt{A_{ISI}}}
\quad ,
\label{cv_gau}
\end{equation}
where $I_0$ is a parameter. For fixed noise variance,
$R \times \sqrt{A_{ISI}}$ should then be
inversely proportional to $\bar I + I_0$ as indeed
verified for $\sigma=100$ in Fig.~\ref{cvisi}.

\subsection{Coexisting coherence resonances}
\label{tre2}

In the silent regime we have encountered two coexisting
coherence resonance phenomena. The first one is related
to the existence of an optimal noise level for the regularization
of the output spike trains and it corresponds to the
effect previously reported in \cite{lee_neiman_1998,yu_wang_2001}.
The second coherence effect can only be detected by considering
the membrane potential dynamics, not being associated
with spikes, but instead with the excitation of quite regular
sequences of subthreshold oscillations occurring
at small noise variances.

\subsubsection{Coherence of the emitted spike trains}

At intermediate noise levels the $P_{ISI}(t)$
reduces its exponential tail and begins to assume the shape
of an inverse
Gaussian distribution (see Fig. \ref{pisi_i6.15}b).
Therefore, the activation mechanism responsible for the firing is gradually
substituted by another stochastic mechanism resembling a
Wiener process with drift.
In correspondence to the transition from one kind of stochastic
process to the other a regularization of the output signal
is observed: this phenomenon is known as {\it
coherence resonance}~\cite{pikovsky_kurths_1997} and has already been
reported for various neuronal models.
Our aim in the present subsection is to give a more detailed
characterization of the CR phenomenon in terms of commonly used indicators
of coherence~\cite{lindner_ojalvo_2004}, like
the coefficient of variation $R$ and the correlation time
$\tau_c$, but also in terms of the saturated conditional entropies
$h_A$, not previously employed in this context.

\begin{figure}[h]
\begin{center}
\includegraphics*[width=7cm]{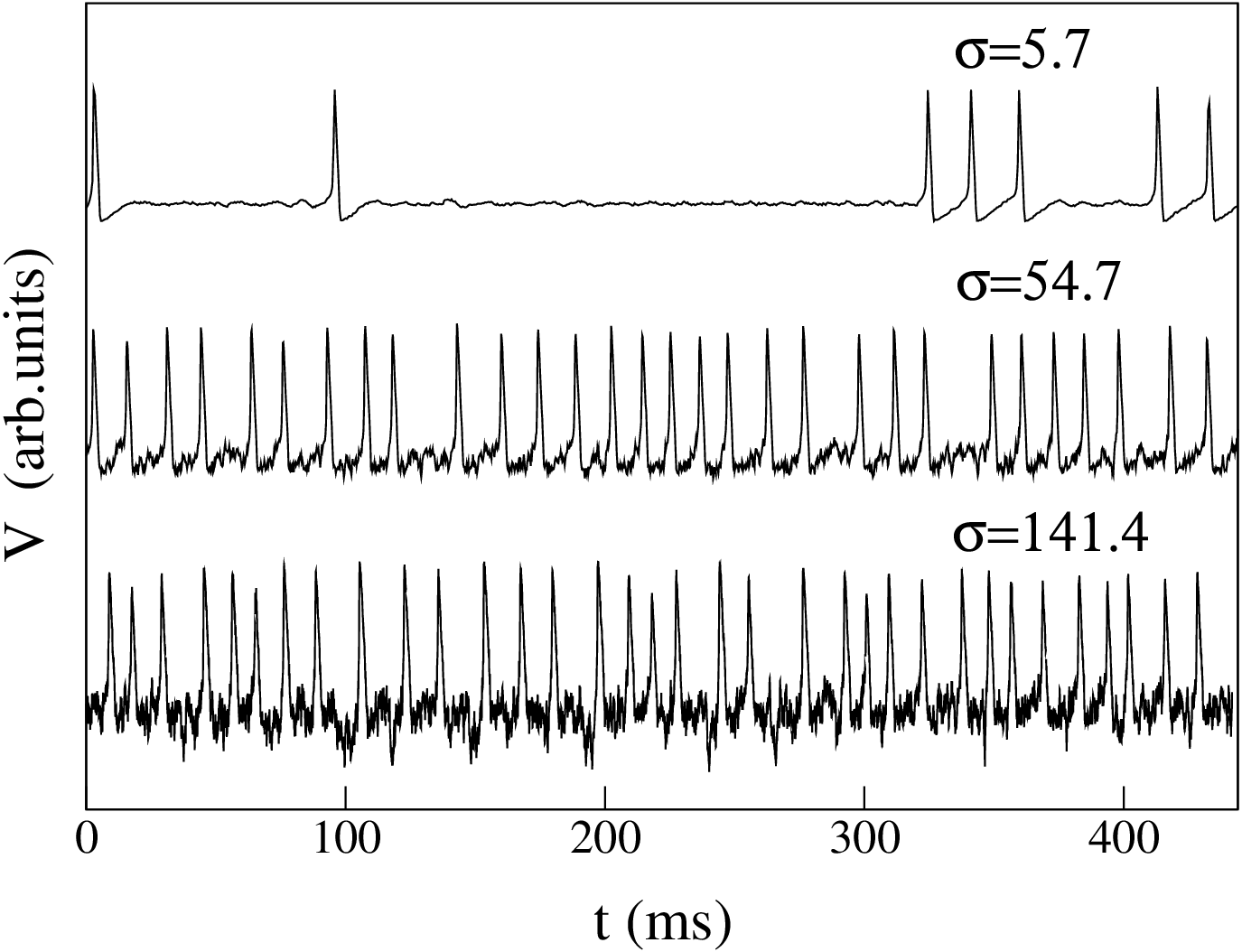}
\end{center}
\caption{Membrane potentials $V(t)$
for various levels of noise $\sigma$ in the
silent regime at $\bar I=5\mu A/cm^{2}$.
}
\label{sig_coh}
\end{figure}

Let us first examine the signals displayed in Fig. \ref{sig_coh}
for $\bar I=5\mu A/cm^{2}$. At low $\sigma$-values one observes rare
spikes induced by the activation mechanism, while for increasing
noise the train becomes more and more regular and finally at high
$\sigma$-values the noise begins to modify even the relative
refractory times.

\begin{figure}[h]
\begin{center}
\includegraphics*[width=6.8cm]{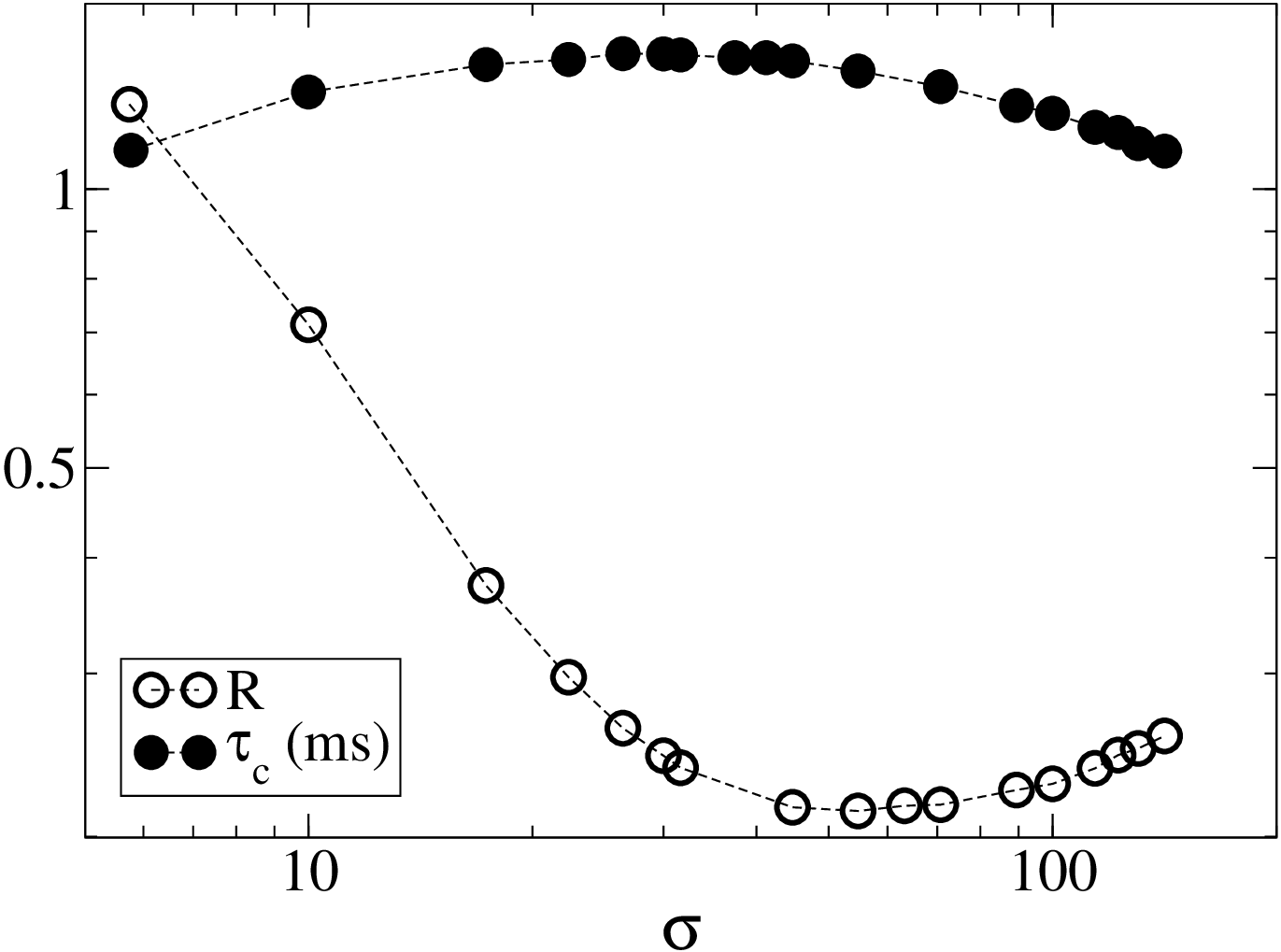}
\end{center}
\caption{$\bar I=5\mu A/cm^{2}$: coefficient of variation
$R$ (empty circles) and correlation time $\tau_c$
(filled circles) as a function of $\sigma$.}
\label{i5_cv_tauc}
\end{figure}

A first quantitative evidence of coherence resonance can be given
in terms of $R$: at low $\sigma$ this quantity tends to one as
expected for Poissonian processes, indeed $R$ reaches a value
slightly greater than one due to the multimodal structure of the
$P_{ISI}(t)$; for increasing $\sigma$ the $S_{ISI}$ decreases
faster than $A_{ISI}$ since the diffusive nature of the process
begins to prevail over the activation effects (exemplified by a
Poissonian distribution for $P_{ISI}(t)$ with $S_{ISI}=A_{ISI}$).
The behavior of $R$ at large $\sigma$-values is given by the
expression (\ref{cv_gau}) and since $A_{ISI}$ slightly decreases
for large noise amplitude the coefficient of variation turns out
to be an increasing function of $\sigma$. A minimum of $R$ is
found at intermediate noise levels due to the predominance of two
different mechanisms at the origin of spiking in two opposite
limits: activated barrier crossing in the limit $\sigma <<
\sigma_c$ and diffusive motion with drift in the opposite limit
$\sigma >> \sigma_c$. Indeed, as shown in Fig.~\ref{i5_cv_tauc} a
minimum is attained for $\sigma \equiv \sigma_{R} \approx 55$ for
$\bar I=5\mu A/cm^{2}$.

A second striking evidence of coherence resonance is given by a
maximum in the correlation time $\tau_c$ estimated directly by
integration of the squared autocorrelation function of the signal
according to Eq. (\ref{tauc}). This maximum occurs for a value smaller than that
obtained for $R$, i.e. $\sigma \equiv \sigma_\tau \approx 30$ for $\bar I=5\mu
A/cm^{2}$.

\begin{figure}[h]
\begin{center}
\includegraphics*[width=7cm]{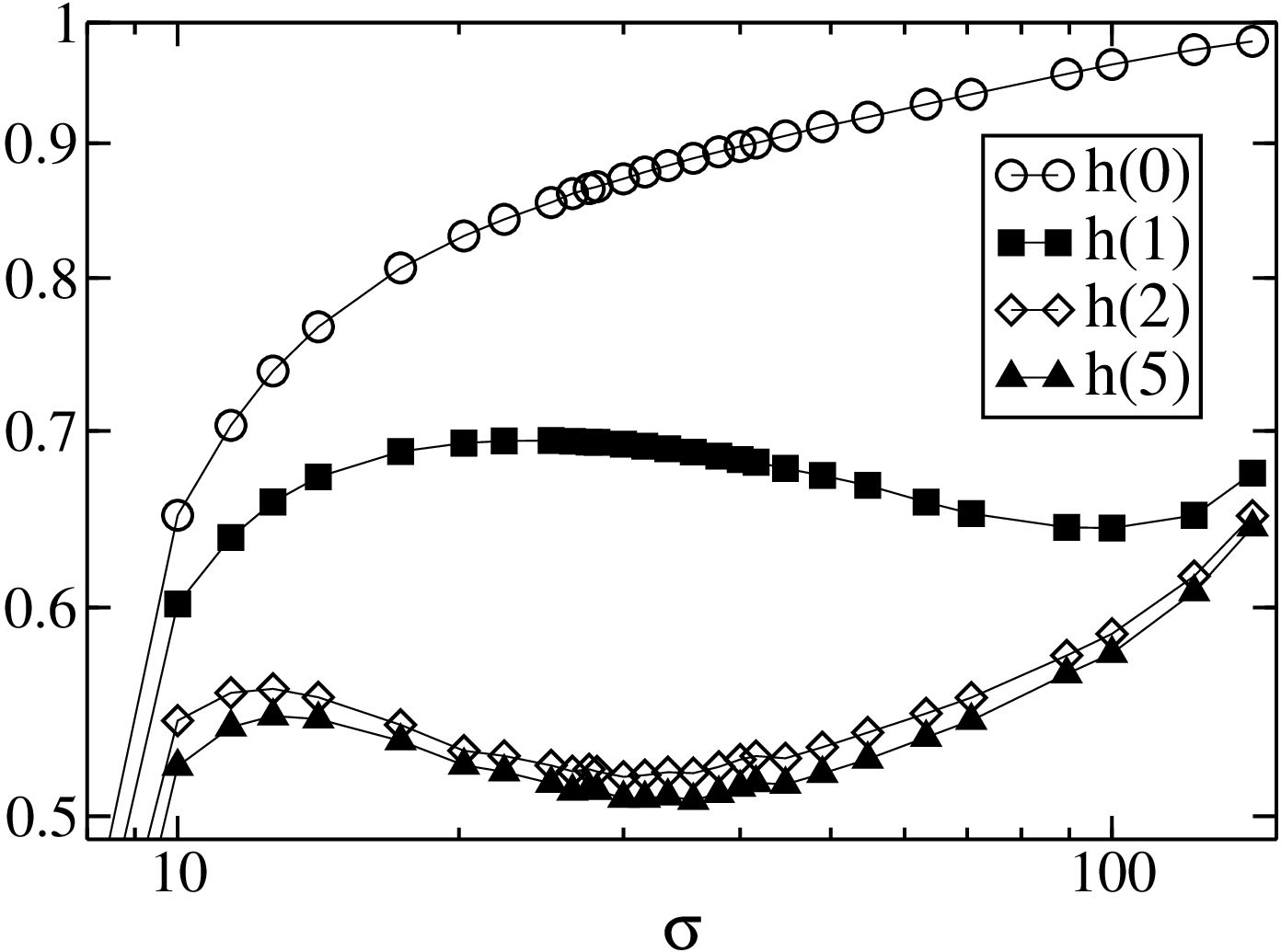}
\end{center}
\caption{Conditional entropies $h(N)$ as a function of $\sigma$
for various word lengths $N$. The data have been obtained for $\bar I=5\mu
A/cm^{2}$ with a resolution of $\Delta t= 5$ ms.}
\label{i5_h_sigma}
\end{figure}

As previously suggested for SR~\cite{neiman_1996}, dynamical
entropies are indicators appropriate to characterize the degree of
unpredictability of the output of a certain nonlinear system,
therefore we expect that they should be useful also in the present
context. In Fig.~\ref{i5_h_sigma} the conditional entropies $h(N)$
are reported for increasing word length $N$ at $\bar I=5\mu
A/cm^{2}$. These quantities saturate to their asymptotical values
$h_A$ for $N = 5$, thus indicating that the present dynamics can
be reproduced in terms of a Markovian process of order $5$ with a
temporal memory $\approx 5 \times \Delta t = 25 ms$. Since for
this current value $A_{ISI}$ varies between 17 ms and 11 ms for
$20 < \sigma < 150$, this means that in order to recover all the
transmitted information it is sufficient to record from two to
three successive spikes. Similar to what has been reported in
\cite{neiman_1996} $h_A$ exhibits a non-monotonic behavior
characterized by a rapid initial growth at small $\sigma$ followed
by a decrease and a minimum at intermediate noise ($\sigma \equiv
\sigma_h \approx 33$). Initially the signal is extremely regular,
being essentially a sequence of zeros in absence of spikes, then
the noise tends to randomize the system and this induces a rapid
increase. Finally, the signal regularizes at finite $\sigma$ and
this leads to the occurrence of a minimum in the asymptotic
entropies.

\begin{figure}[h]
\begin{center}
\includegraphics*[width=7cm]{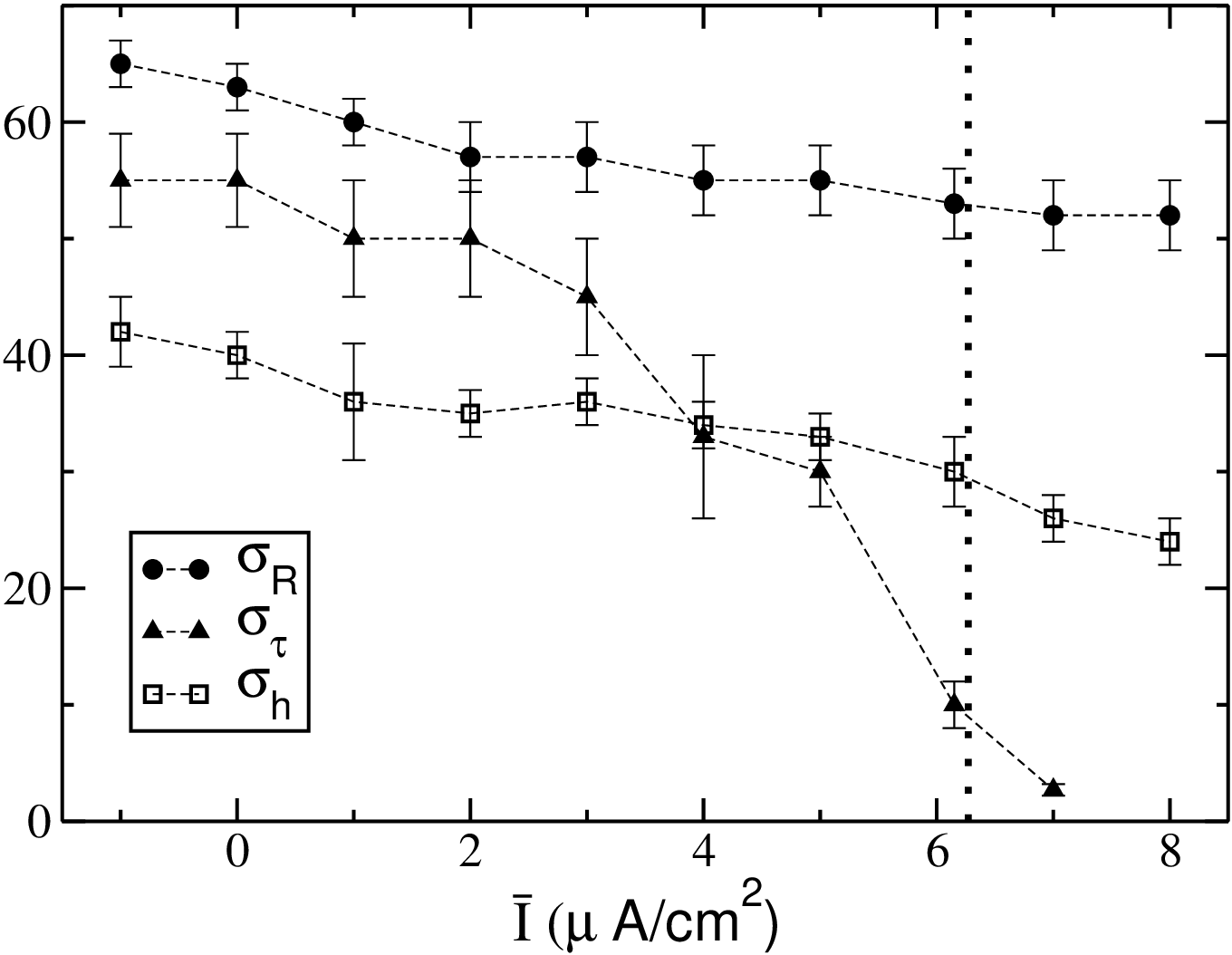}
\end{center}
\caption{Optimal noise standard deviations
corresponding to coherence resonance
versus $\bar I$ for the three considered indicators:
namely, $\sigma_{R}$ (filled circles),
$\sigma_\tau$ (filled triangles)  and
$\sigma_h$ (empty squares).
The vertical dotted line indicates $I_{SN}$,
while the dashed lines are guides for the eyes.
}
\label{respos}
\end{figure}

We can safely affirm tha the three employed indicators agree in
indicating a clear coherence resonance effect at the considered
current $\bar I=5\mu A/cm^{2}$. Furthermore, we performed a
similar analysis at various current $\bar I$ in the silent and in
the bistable regime and the corresponding values $\sigma_{R}$,
$\sigma_h$, $\sigma_\tau$ are shown in Fig.~\ref{respos}. For all
indicators the optimal amount of noise needed to observe CR
decreases for increasing current. However, while $\sigma_{R}$ and
$\sigma_{h}$ seem not to vary dramatically in the examined range,
$\sigma_\tau$ decreases noticeably approaching the SN bifurcation.
This behavior seems to be due to the fact that the signal
autocorrelation function does not only register coherence effects
associated to the ISIs, but also those related to the subthreshold
oscillations occurring at quite low noise (as reported in the next
subsection). The interaction between these two resonances leads to
the enhanced decrease of $\sigma_\tau$, somehow the ISI resonance
is entrained by the second one and shifted towards small $\sigma$.
Other indications that the signal becomes more and more correlated
approaching $I_{SN}$ are given by the corresponding increase of
the maximal correlation time and by the decrease of the minimum of
$R$ and of $h_A$ (all measured at the resonance). This scenario is
consistent with the fact that the activation barrier $W_S$
separating the rest state from the excited state also decreases
with $\bar I$ and therefore the latter becomes more accessible at
lower noise variances.

\subsubsection{Coherence of the subthreshold oscillations}

\begin{figure}[h]
\begin{center}
\includegraphics*[width=7cm]{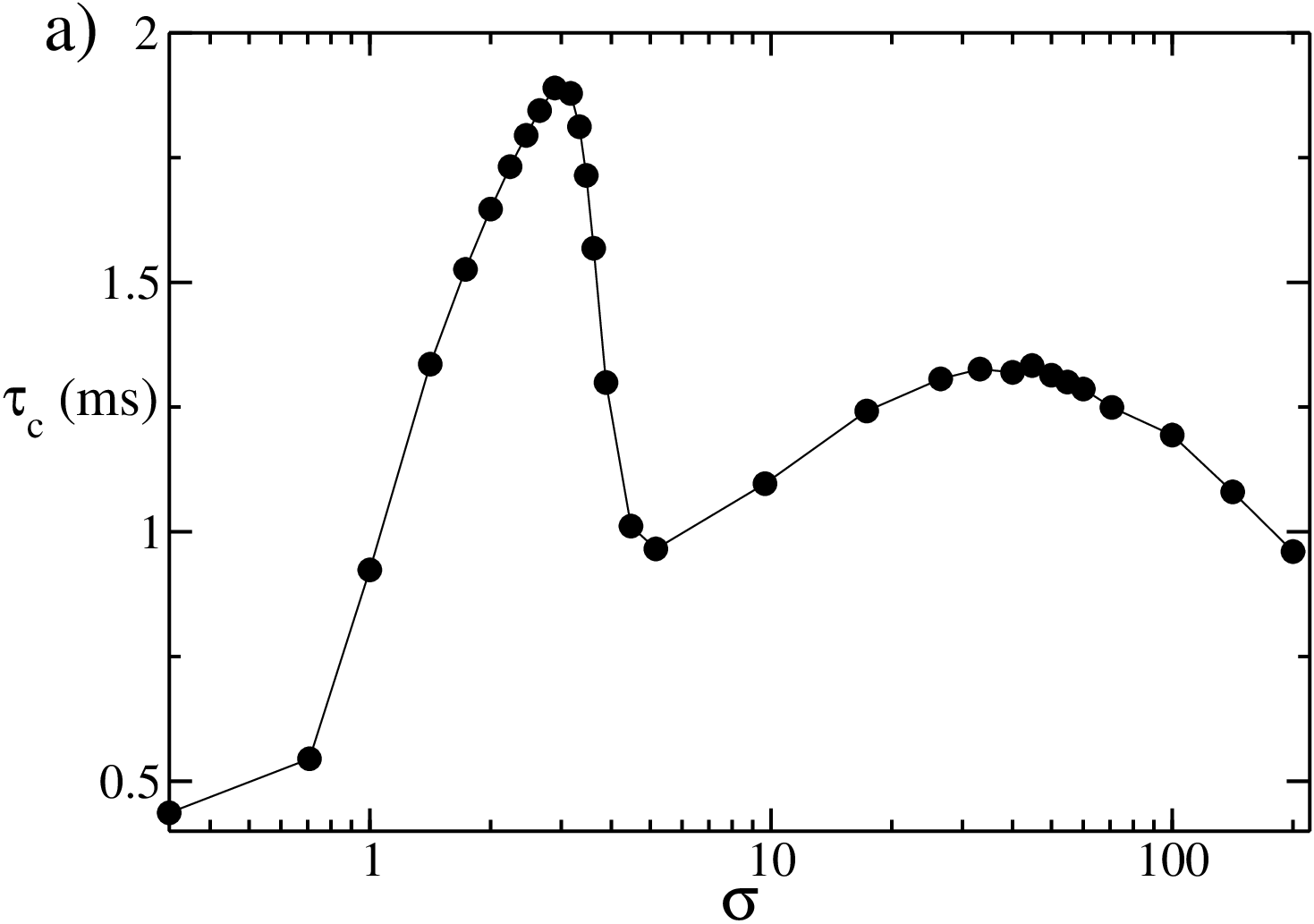}
\includegraphics*[width=7cm]{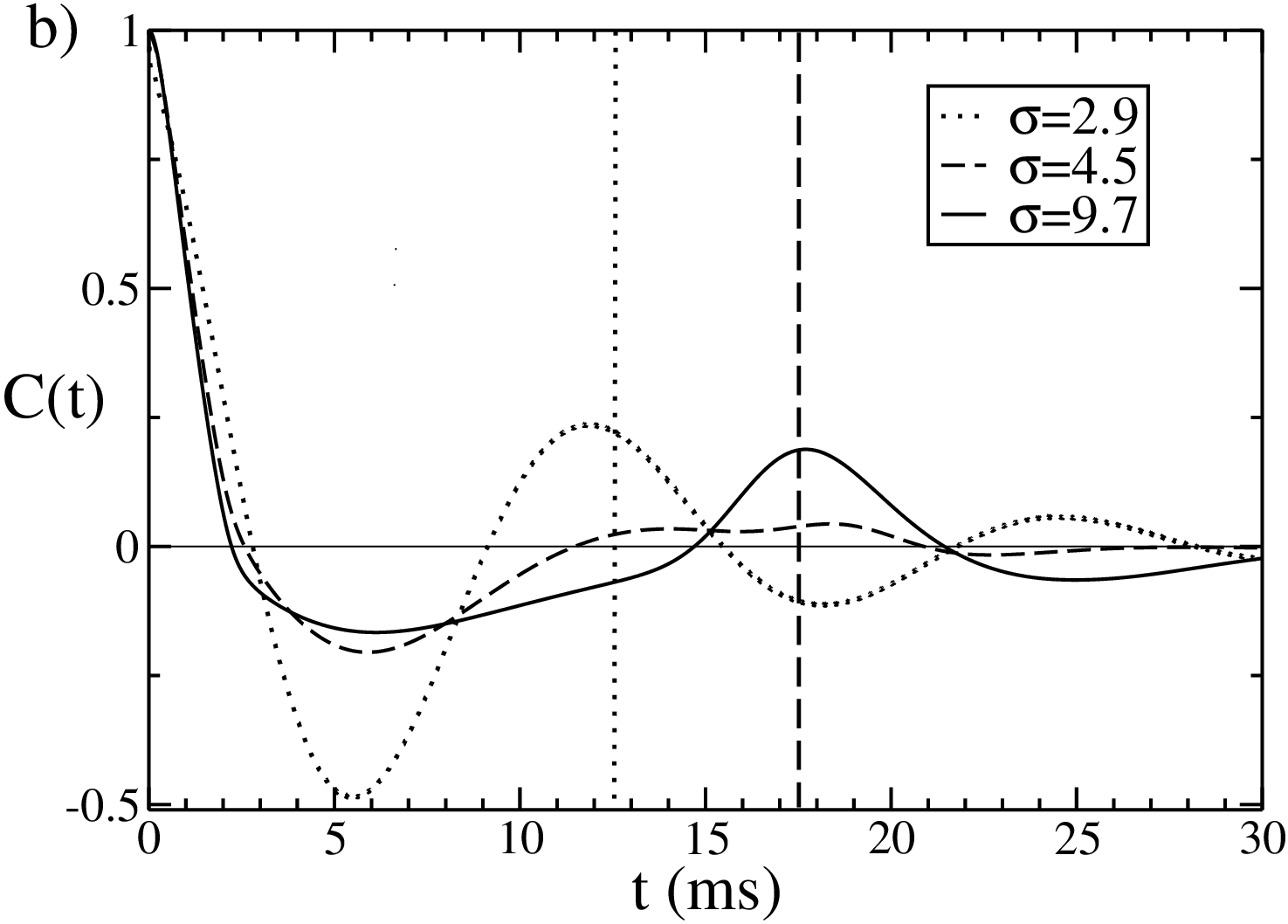}
\end{center}
\caption{$\bar I= 4 \mu A/cm^{2}$: (a) $\tau_c$ (filled circles) as a function of $\sigma$,
the solid line is a guide for the eyes; (b) autocorrelation functions
$C(t)$ for three different noise values (namely $\sigma=2.9,4.5$ and
9.7), the vertical dotted
line indicates the period $T_l$, while the vertical dashed
line refers to $t^{(1)}_{ISI}$ for $\sigma=9.7$.}
\label{tauI4}
\end{figure}

 Let us now discuss in more detail the origin of the second coherence
 resonance observed at very low noise variance.
 In Fig.~\ref{tauI4} the behavior of the correlation time $\tau_c$
 as a function of $\sigma$ is shown for $\bar I= 4 \mu A/cm^{2}$
 in a wider range of noise with respect to the data shown in Fig.~\ref{i5_cv_tauc}.
 In that figure the examined noise range was restricted to values
 for which the statistics of emitted spikes was sufficiently rich to ensure
 a meaningful definition of $R$.
 In Fig.~\ref{tauI4}a $\tau_c$ reveals two clear maxima:
 the higher one located at $\sigma \approx 3$ and
 the lower one at $\sigma \approx 33$. The origin of the first peak
 can be understood as follows: for $\sigma < 3$ almost no spikes
 are emitted by the neuron, however, the increase
 of noise tends to stimulate series of subthreshold oscillations that are
 more and more correlated; for $\sigma > 3$ the statistics of the emitted
 spikes begins to be no more negligible and the occurrence of
 rare spikes tends to decorrelate the signal leading to a
 decrease of $\tau_c$. By further increasing the noise variance
 the signal begins to be characterized by sequences of spikes, therefore
 the autocorrelation function starts to register essentially the
 correlation of these events and $\tau_c$ reveals a second peak
 related to the regularization of the spike trains, this is the coherence
 phenomenon previously discussed.

 The transition from one dynamical regime to the other can be
 better understood by examining the autocorrelation function of $V(t)$
 for various values of $\sigma$. As shown in Fig.~\ref{tauI4}b,
 for $\sigma \approx 3$ the autocorrelation function $C(t)$ reveals oscillations
 of period $\approx T_l$, while at $\sigma \approx 9.7$ the maxima of
 $C(t)$ are located in correspondence with multiples of $t \approx t^{(1)}_{ISI}$.
 In between these two values there is a transition from an output signal dominated by
 the subthreshold oscillations to a signal characterized by trains of
 spikes, this transition can be located at $\sigma \approx 4.5$, since at this value
 $C(t)$ exhibits, at the first oscillation, two maxima one located at
 $t \approx T_l$ and one at $t \approx t^{(1)}_{ISI}$.

 To summarize, for the HH model in the silent regime
 two kind of coherence resonances can be observed, one at quite low noise related
 to the excitation of subthreshold oscillations around the rest state and another one
 at higher noise due to the regularization of successive $ISI$s associated to the spikes
 emitted by the neuron. Obviously, since the first resonance is not related to
 spike occurrence it cannot be revealed by $R$ or by $h$, but only by
 $\tau_c$.

\section{Bistable regime}
\label{quattro}

\begin{figure}[h]
\begin{center}
\includegraphics*[width=9cm]{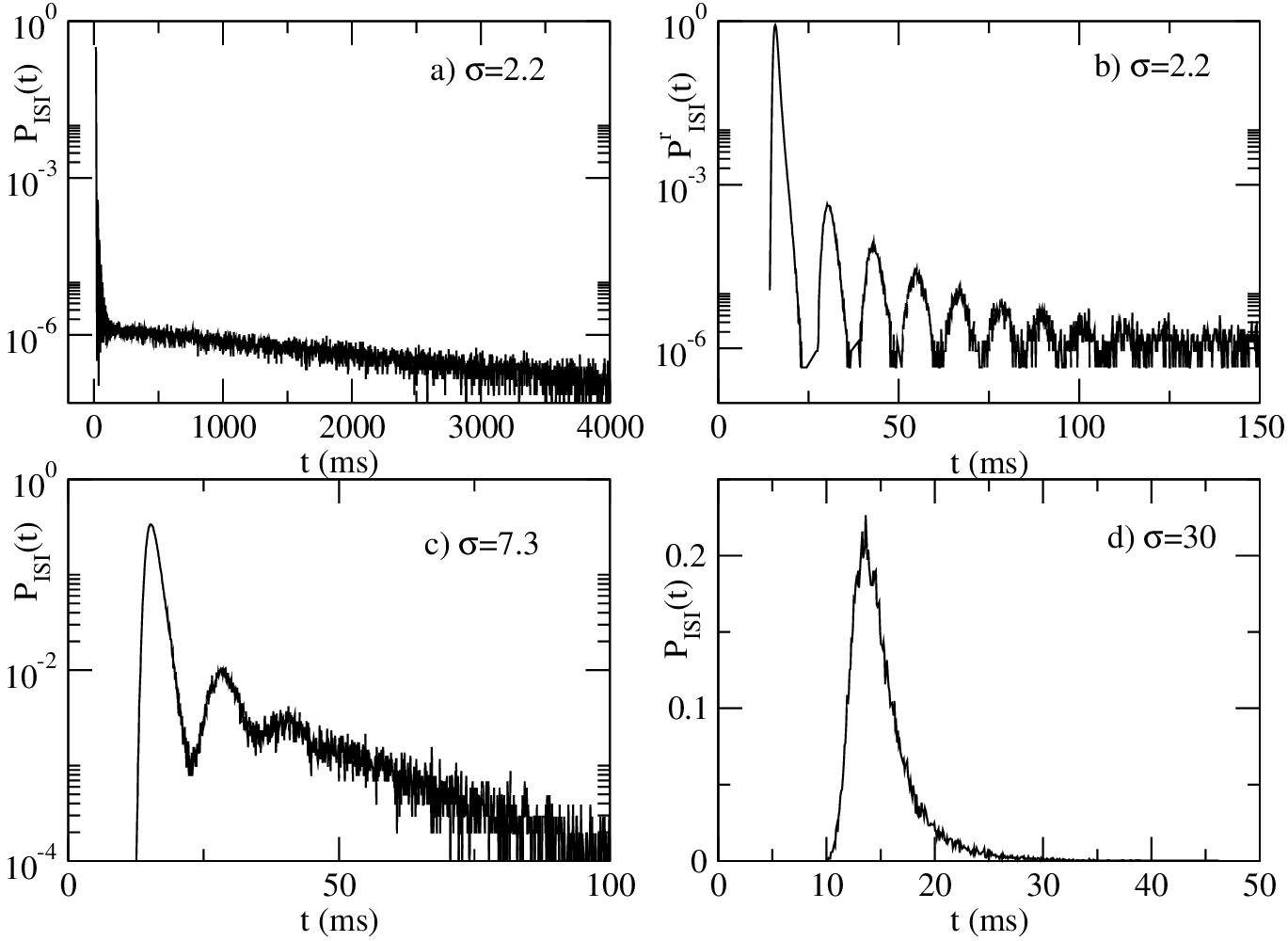}
\end{center}
\caption{$P_{ISI}(t)$ in the bistable regime at $\bar I=8\mu
A/cm^{2}$. $P_{ISI}(t)$ is displayed for various values of the
noise: $\sigma=2.2$ (a), $\sigma=7.3$ (c), and $\sigma=30$ (d).
Additionally in (b) $P^r_{ISI}(t)$ is shown for $\sigma=2.2$.}
\label{i8_histo}
\end{figure}

\begin{figure}[h]
\begin{center}
\includegraphics*[width=8cm]{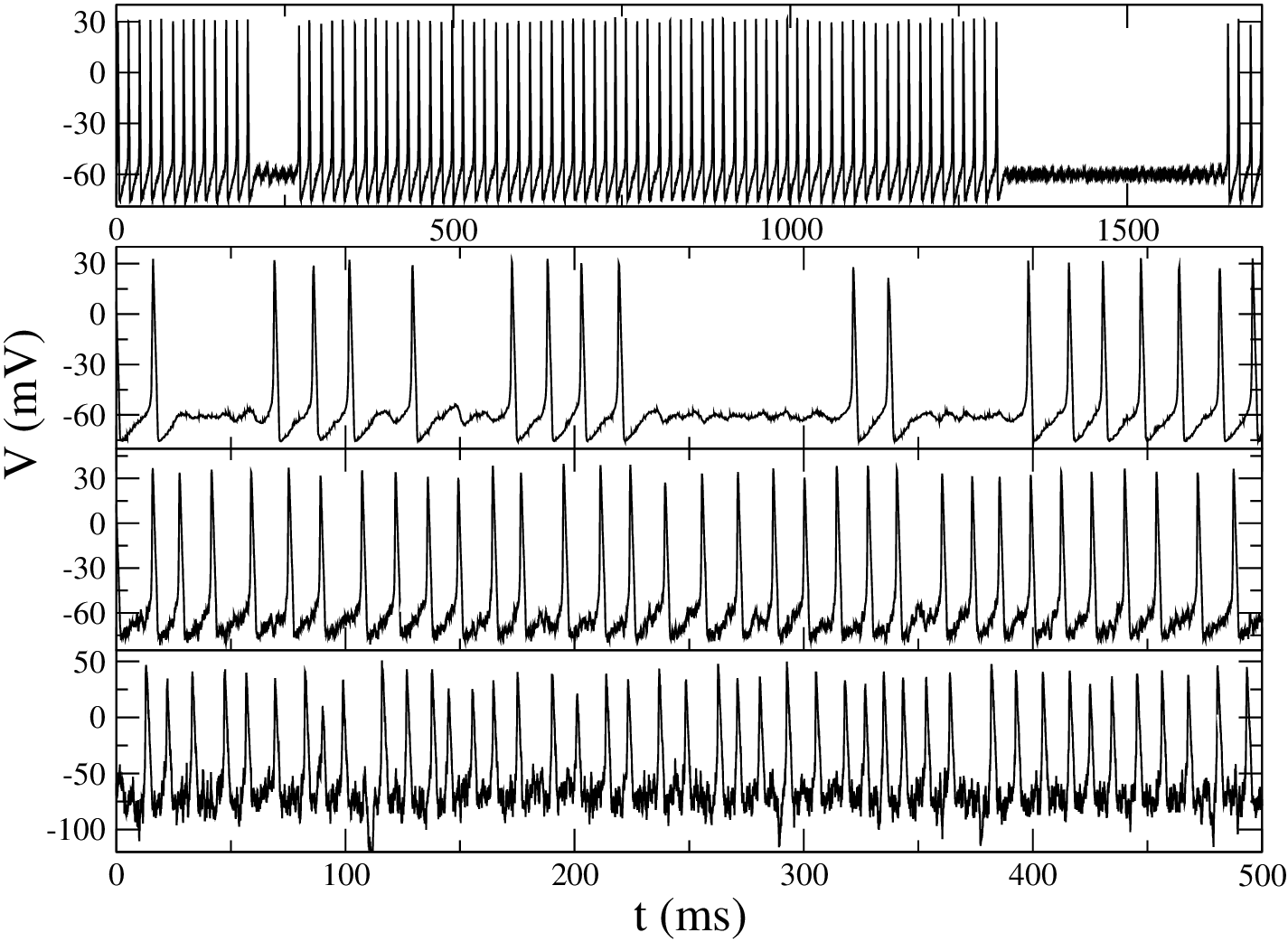}
\end{center}
\caption{Membrane potential $V(t)$ evolution in time for
various value of the noise variance. The top signal
refers to $I = 8\mu A/cm^{2}$ and $\sigma=3.2$, the other
three curves correspond (from top to bottom) to
$\sigma= 6.8$, 38.7 and 141.6 and
$I = 7\mu A/cm^{2}$.
}
\label{i8_segnale}
\end{figure}

\begin{figure}[h]
\begin{center}
\includegraphics*[width=8cm]{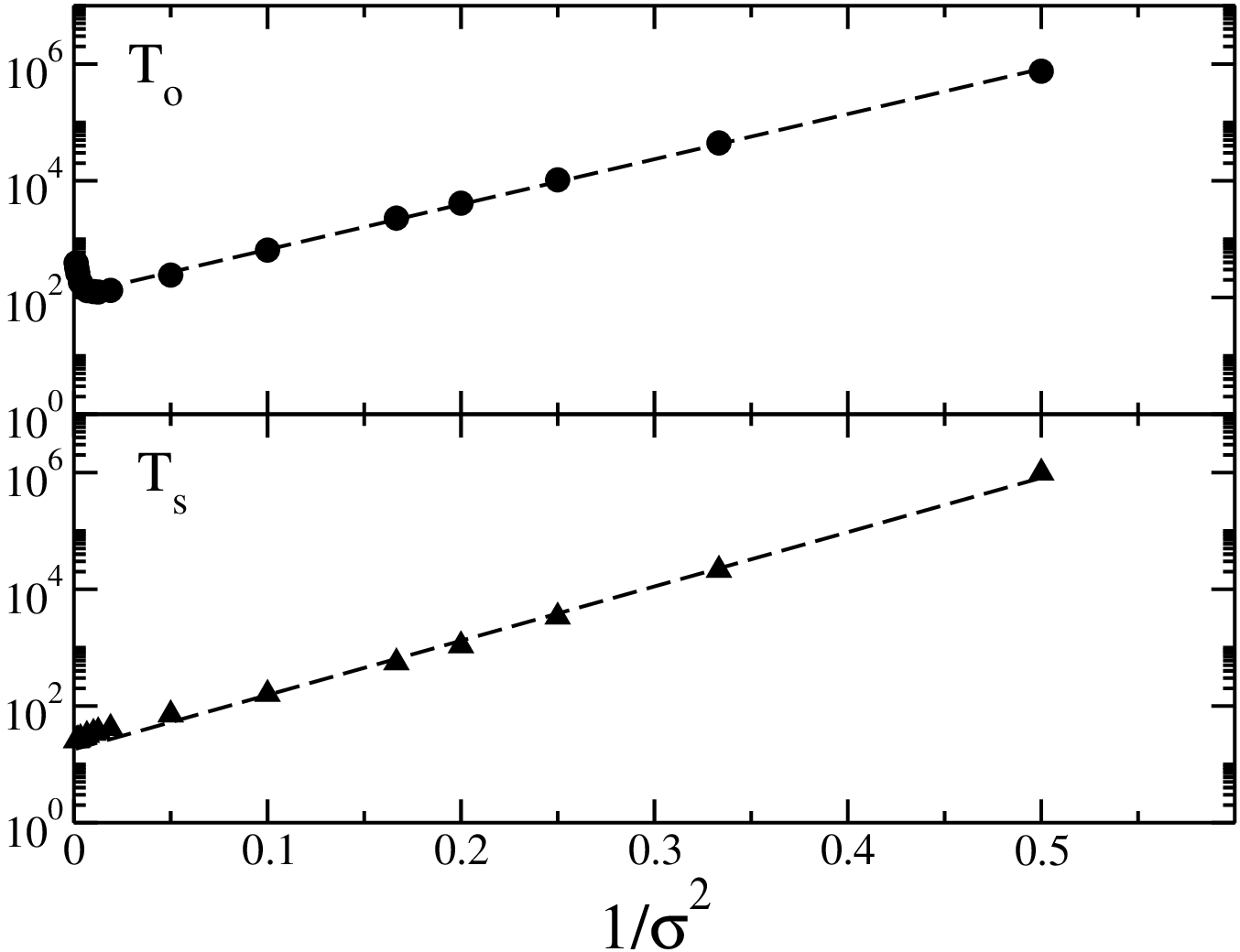}
\end{center}
\caption{Residence times $T_O$ and $T_S$ as a function of
$1/\sigma^2$ for low noise variance values: the dashed
lines are fits to the numerical data with
expression (\ref{activation_times}) for $\sigma \le 4$,
the corresponding barriers are $W_S=21.4\pm0.5$ and $W_O=17.9\pm0.3$.
All data refer to $\bar I=8\mu A/cm^{2}$.
}
\label{i8_analisitempidiresidenza}
\end{figure}

In the bistable regime the modifications of the shape of the ISI
distributions due to the effect of noise resemble those found in
the silent regime, apart for very low noise variances (see Fig.
\ref{i8_histo}). As shown in Fig. \ref{i8_histo}a, in this latter
case one observes a quite pronounced peak corresponding to
$1/{\nu_c}$ followed by smaller multi-peaks and an exponential
tail. This reflects the fact that the dynamics due to the noise
switches back and forth between the two coexistent states:
oscillatory and silent. (cf. the time evolution of the membrane
potential reported in Fig.~\ref{i8_segnale}, in particular at
$\sigma=3.2$ (for $\bar I=8\mu A/cm^{2}$)). This dynamical
behavior is reminiscent of the motion of one particle in a double
well subjected to thermal fluctuations. Therefore, the residence
times in the two states should be related with the noise amplitude
by a Kramers relationship in the standard manner
\begin{equation}
T_x \propto {\rm e}^{W_x/\sigma^2}
\quad ,
\label{activation_times}
\end{equation}
where $x=S$ for the silent state and $x=O$ for the oscillatory
state. Moreover, the distributions of $T_S$ (resp. $T_O$)
should be Poissonian due to the stochastic nature of the jumping
from one state to the other. In order to verify Eq.
(\ref{activation_times}) we have estimated the times $T_S$ and
$T_O$ as a function of $\sigma$ for small values of this parameter.
The data, displayed in Fig. \ref{i8_analisitempidiresidenza},
confirm that in the low noise limit both times can be described in
terms of an activation process induced by noise fluctuations.
Moreover, barriers $W_O$ and $W_S$ have been estimated for
various currents and are reported in Fig. \ref{barrier}.
For $ I_{SN} < \bar I < 8\mu A/cm^{2}$
the most stable state is the fixed point, while above
$I = 8\mu A/cm^{2}$ the stability of the oscillatory state prevails.

\begin{figure}[h]
\begin{center}
\includegraphics*[width=7cm]{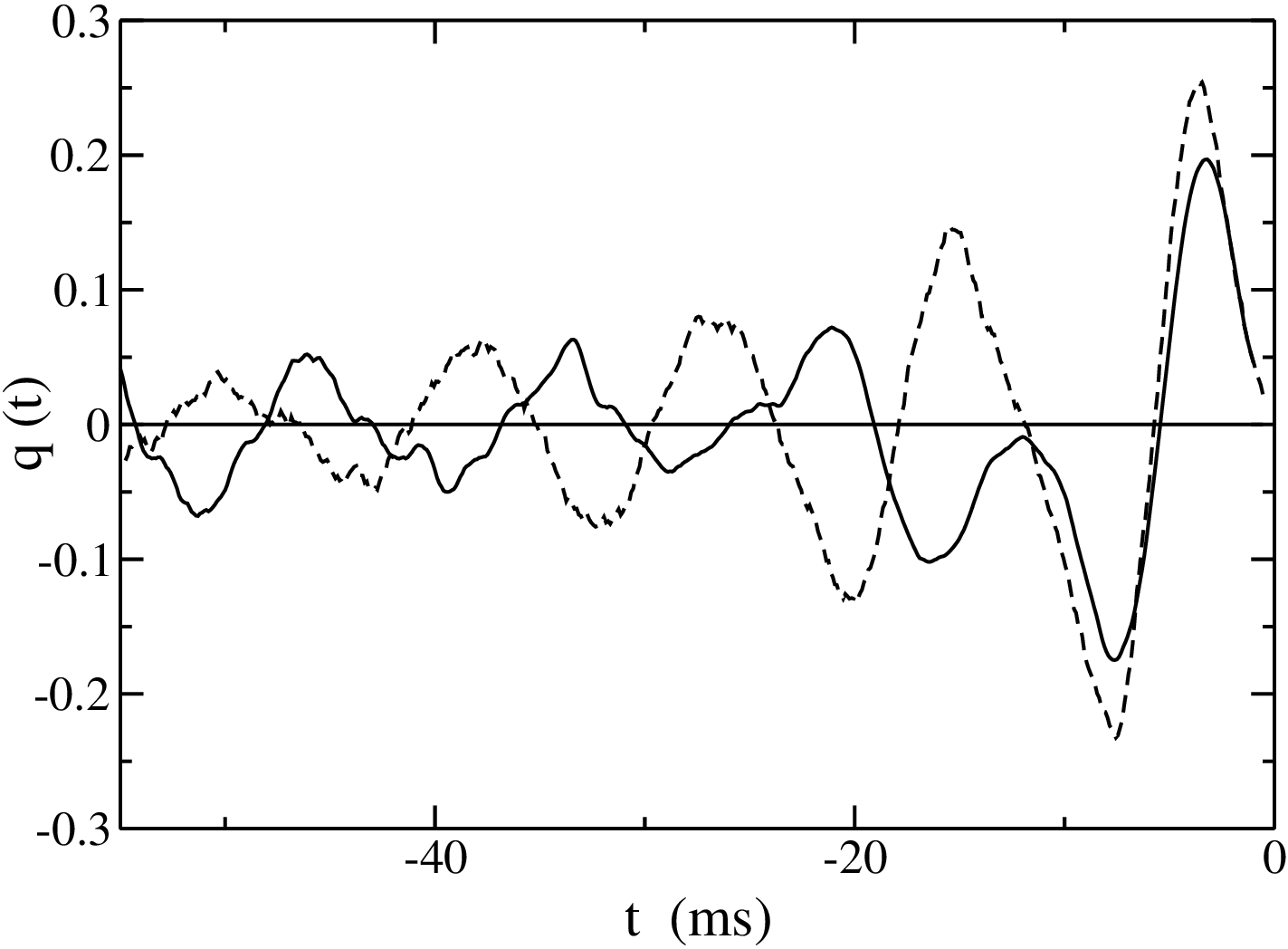}
\end{center}
\caption{STAF
for long ISIs $q_L(t)$ (dashed line)
and intermediate ISIs $q_I(t)$ (solid line) for
$\bar I=8\mu A/cm^{2}$ and $\sigma= 3.2$.
$q_L(t)$ has been estimated by averaging over
$ISI > 70$ ms, while $q_I(t)$ refer to
$25 ms < ISI < 70$ ms.}
\label{spc_i8}
\end{figure}

Let us now come back to the $P_{ISI}(t)$ for $\sigma=2.2$, in this
case we have estimated the distributions of the ISIs during the
oscillatory state only, let us term this probability distribution
as $P^r_{ISI}(t)$. This distribution, reported in Fig.
\ref{i8_histo}b, reveals a multimodal structure. While the first
peak corresponds to $ \approx 1/\nu_c$, the other peaks are
related to the linear damped oscillations towards the fixed point,
i.e. $t^{(k+1)}_{ISI}-t^{(k)}_{ISI} \approx k T_l$ for $k \ge 1$.
The origin of this multimodal structure can be understood by
comparing the STAF $q_I(t)$ preceding the emission of a spike for
intermediate ISI-values (namely for ISI durations corresponding to
the second and third peak of the distribution) with the STAF
$q_L(t)$ associated with long ISIs falling in the exponential time
tail. Analogously to what observed in the silent regime $q_L(t)$
reveals oscillations of period $ \approx T_l$, indicating that the
neuron will return to fire, once entered in the silent state,
mainly when stimulated via current oscillations of the proper
period and in phase with $V(t)$ oscillations (see Fig.~\ref{pot}).
As shown in Fig. \ref{spc_i8}, $q_I(t)$ has a quite peculiar
behavior, it also exhibits oscillations of period $\approx T_l$,
but these oscillations are in anti-phase with respect to those of
$q_L(t)$, apart from the last two preceding the firing of the
neuron. This means that, once entered in the oscillatory state,
the neuron can become silent only if a series of current
oscillations (in anti-phase with respect to the damped
oscillations of $V(t)$ following a spike) inhibit repetitively
spike emission allowing a sufficient decrease of $V(t)$ leading
the system in the attraction basin of the rest state. If by chance
one of these oscillations returns in phase with $V(t)$ (during the
relaxation) the neuron will fire immediately after (as shown in
Fig. \ref{spc_i8}). This is the mechanism at the origin of the
various peaks present in $P_{ISI}(t)$ at low noise values, some
indications consistent with this scenario have been recently
reported in Ref.~\cite{aguera03}.

\begin{figure}[h]
\begin{center}
\includegraphics*[width=9cm]{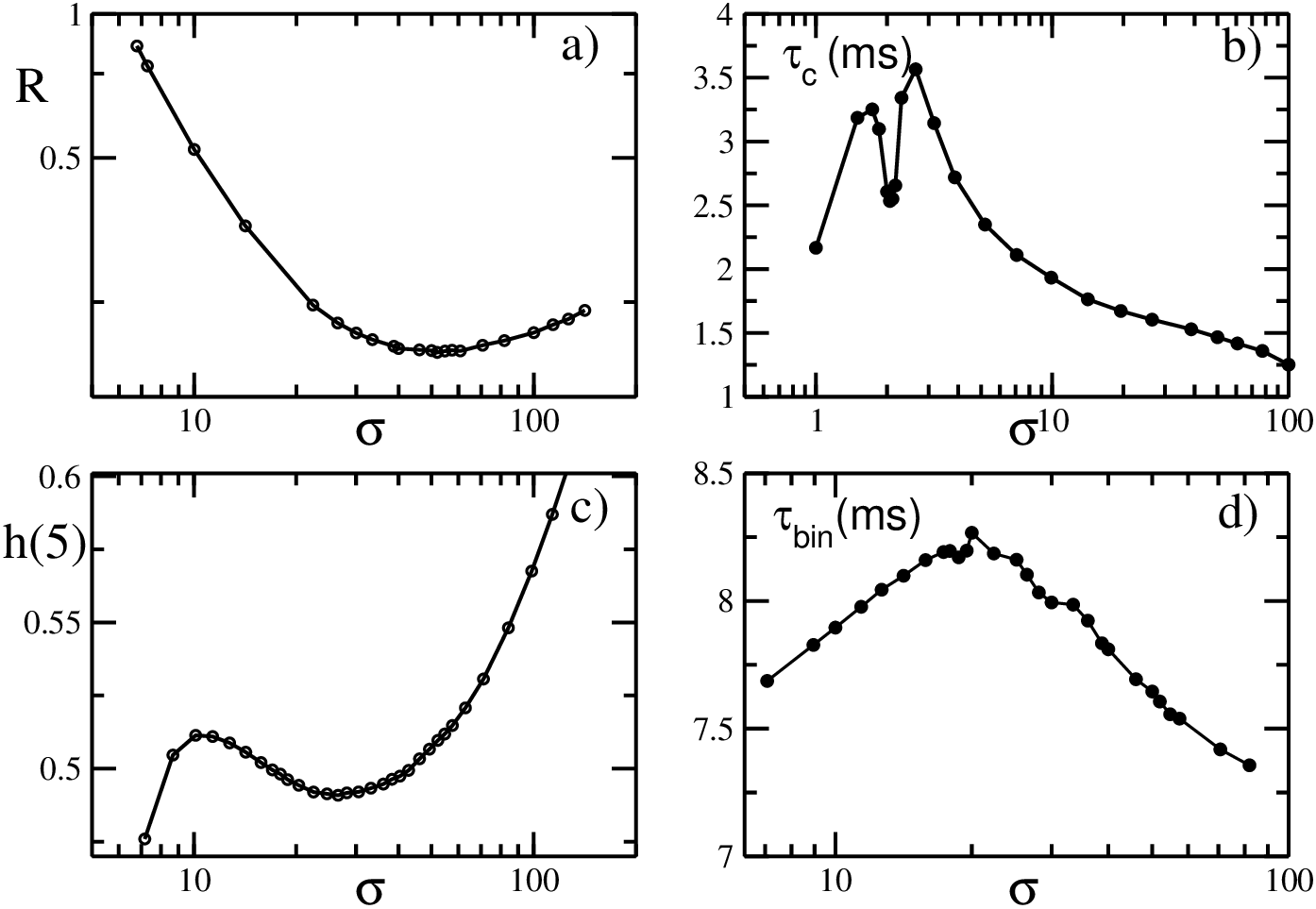}
\end{center}
\caption{Regime of bistability ($\bar I=7\mu A/cm^{2}$): R (a)
, $\tau_{c}$ (b), $h_A=h(5)$ (c) and $\tau_{bin}$ (d) as a
function of $\sigma$.}
\label{i8_sigma_cv_tauc}
\end{figure}

As far as CR is concerned, a
clear resonance is observable by inspecting
the dependence of $R$ and the saturated
$h_A$ as a function of noise intensity for $\bar I=7\mu A/cm^{2}$
(see Fig.  \ref{i8_sigma_cv_tauc}).
On the other hand, $\tau_c$ exhibits two almost coinciding
maxima at very low noise: one due to subthreshold oscillations
at $\sigma=1.8$ and the other one related to ISI coherence
at $\sigma_\tau=2.6$. At higher currents the two maxima merge
and they are no more distinguishable one from the other.
In order to
understand if a ISI coherence effect is still clearly
discernible at the level of correlation functions
we have estimated the time $\tau_{bin}$ defined in terms of
$C_{bin}$ (\ref{corr_bin}).
This quantity, which is not influenced by the subthreshold
oscillations, shows a clear maximum at $\sigma = 20$
(see Fig.  \ref{i8_sigma_cv_tauc}d). Since $C_{bin}$
and $h_A$ measure both dynamical features of the ISI
sequence we expect that they should give similar information,
indeed for this current $\sigma_h = 26 \pm 2 $.
These results seem to indicate that the
information reduction needed to pass from the signal to the ISIs
allows to better single out the ``standard'' coherence phenomenon in this
regime. By further increasing the current the only indicator that
continues to signal a coherence resonance effect is $R$, while
at $\bar I=9\mu A/cm^{2}$ the minimum present in $h$ becomes a shoulder
and also $\tau_{bin}$ does not exhibit maxima at intermediate noise.
The main difference between
the dynamics at $\bar I=7\mu A/cm^{2}$ and $\bar I=9\mu A/cm^{2}$ is
related to the fact that the oscillatory state becomes more stable
than the silent state at this latter current (i.e. $W_O > W_S$).
To summarize, the coexistence of two coherence resonance is observed
also in the bistable regime until the fixed point remains the
most stable solution (i.e. for $\bar I \le 8\mu A/cm^{2}$).

\section{Repetitive firing regime}
\label{cinque}

\begin{figure}[h]
\begin{center}
\includegraphics*[width=9cm]{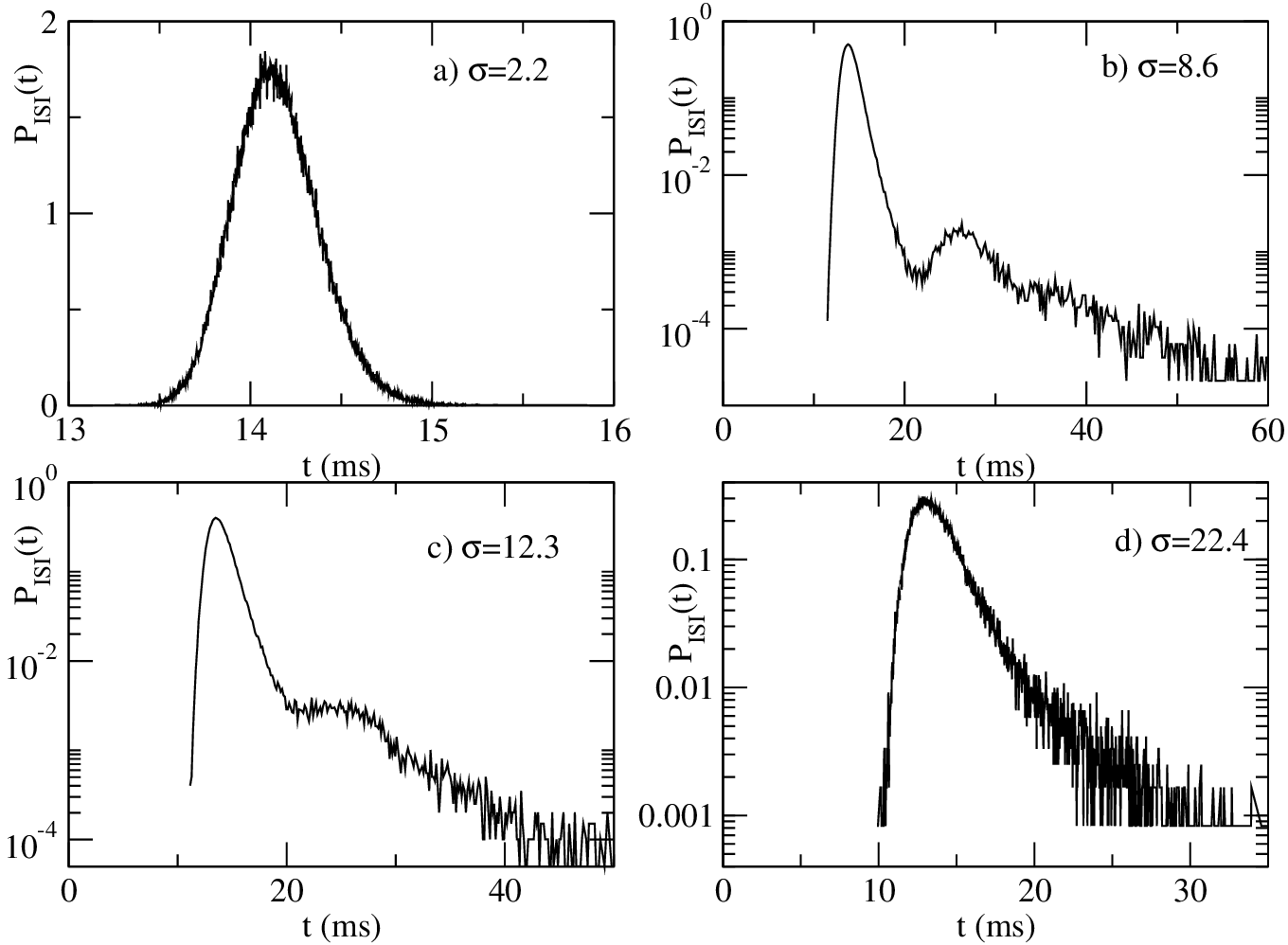}
\end{center}
\caption{$P_{ISI}(t)$ corresponding to $\bar I=11\mu A/cm^{2}$ for
various $\sigma$-values: $\sigma=2.2$ (a), 8.6 (b), 12.3 (c)
and 22.4 (d).} \label{i11_isto}
\end{figure}

Above $I_{HB}$ the system is in a stable regime of repetitive
firing if subjected to a constant current. As we can see from
Fig.~\ref{i11_isto}, the noise has different
effect this time. In particular, for sufficiently small $\sigma$
we observe that $P_{ISI}(t)$ is essentially a
Gaussian centered around the repetitive firing period $1/\nu_c$
(see Fig.~\ref{i11_isto}a). Initially for $1 < \sigma < 3$,
$S_{ISI}$ increases
linearly with $\sigma$, as expected for additive noise of sufficiently small intensity.
Upon further increasing the variance of the noise, the distribution
becomes again multimodal (as shown in Fig. \ref{i11_isto}b) with
peaks located at integer multiples of $1/\nu_c$, but this time
the activation tail is almost absent.
The multi-peak structure (mainly limited to two peaks only)
is due to the fact that
sometimes the neuron fails to emit a spike at $1/\nu_c$ because of
a current fluctuation in anti-phase with the suprathreshold
oscillations of the membrane potential.
This mechanism is analogous to the one already described in the previous
Section to explain the multi-peaks observed in the bistable regime.
At larger $\sigma$ the second peak reduces to a shoulder
(see Fig.~\ref{i11_isto}c) and then $P_{ISI}(t)$ converges towards
an inverse Gaussian.

\begin{figure}[h]
\begin{center}
\includegraphics*[width=8cm]{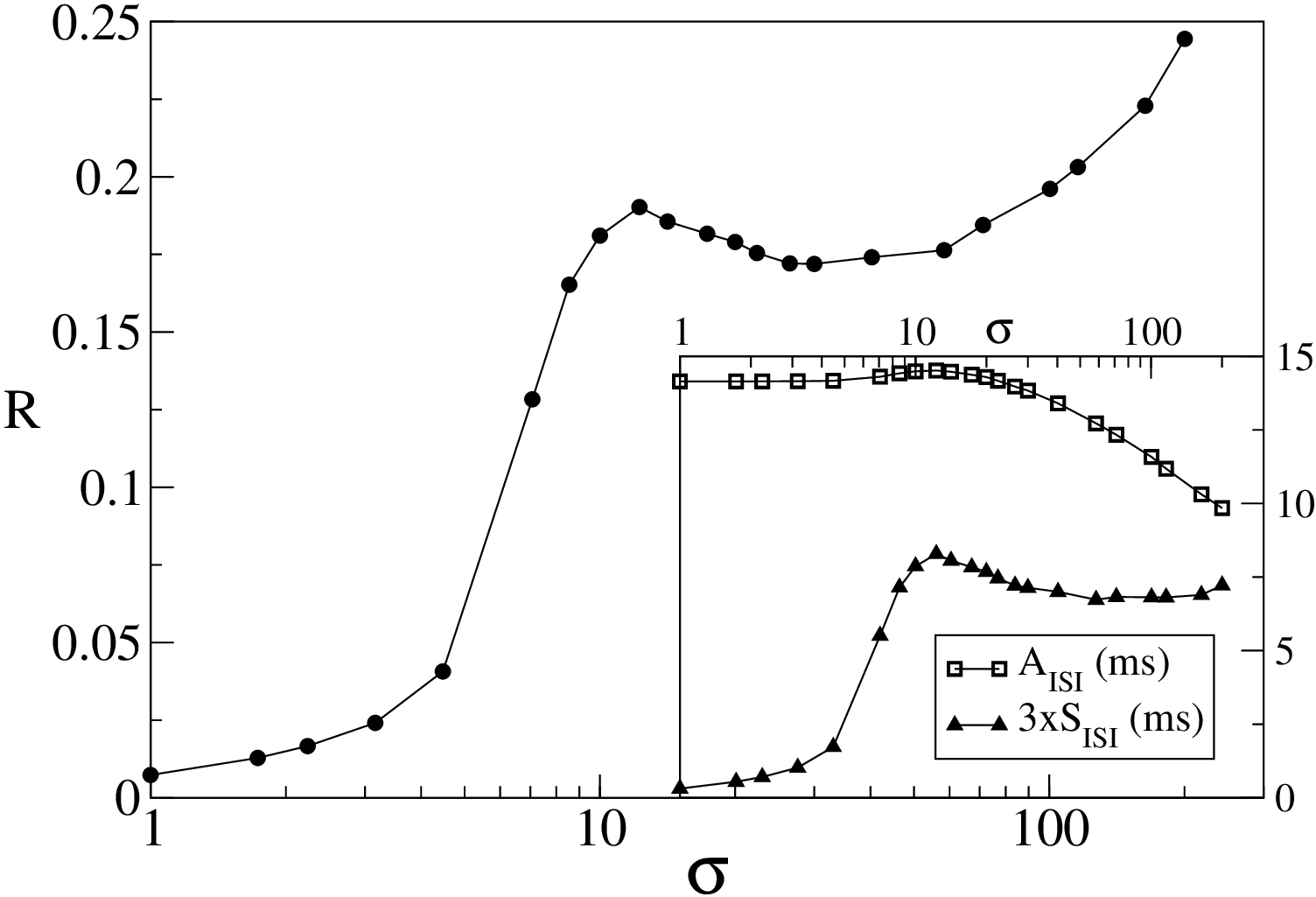}
\end{center}
\caption{$R$ (filled circles) versus $\sigma$
for $\bar I=11\mu A/cm^{2}$. In the inset are reported
$A_{ISI}$ (empty squares) and $S_{ISI}$ (filled triangles) for the
same current.} \label{i11_cv}
\end{figure}

In this regime the only indicator giving evidence of CR is the
coefficient of variation $R$, while for $\sigma \to 0$ the
correlation time $\tau_c$ diverges to infinite, since the system
converge to a stable limit cycle, and the conditional entropies do
not exhibit any relative minimum. As shown in Fig.~\ref{i11_cv},
$R$ has a maximum at $\sigma=12.3$ (for $R \approx 0.19$) followed
by a minimum at $\sigma=30$ (corresponding to $R \approx 0.17$).
However, due to the very limited variation of $R$, it is difficult
to appreciate from the signal evolution some difference between
$\sigma=12.3$ and 30. A similar effect of maximal spike train
incoherence was observed for a leaky integrate-and-fire model with
an absolute refractory period for supra-threshold base
current~\cite{Lindner02}. In that case a maximum occurs at
intermediate noise values since perfectly regular spiking was
found for vanishing noise and in the large noise limit. In our
model the maximum in $R$ is related to the emergence of the
multi-peak structure of $P_{ISI}(t)$. In the limit $\sigma \to 0$
we observe regular spiking (i.e. $R \to 0$), the introduction of
noise in the system initially leads to small irregularities in the
spiking (reflected in the linear increase of $S_{ISI}$ with
$\sigma$) and then to the emergence of the second peak in
$P_{ISI}(t)$, that induces an abrupt growth in the $S_{ISI}$-value
(as shown in the inset of Fig.~\ref{i11_cv}) and also in $R$. At
higher noise levels ($\sigma > 23$) the dynamics reduces
essentially to a Wiener process plus drift, this implies a merging
of the two peaks associated with a simultaneous decrease of
$S_{ISI}$ followed by a saturation. The further randomization of
the dynamics leads to a decrease of $A_{ISI}$, that tends to its
asymptotic value, which is the refractory time. In addition, this
behavior of $A_{ISI}$ is responsible for the minimum of $R$ at
$\sigma=30$ and for the successive growth consistent with Eq.
(\ref{cv_gau}).

\section{Summary and Final Remarks}
\label{sei}

In the present paper we have studied the Hodgkin-Huxley model in
the high-input regime subjected to stochastic trains of
uncorrelated inhibitory and excitatory kicks. Our analysis
suggests that the specific shape of the time distribution of the
arrival times play no role in this framework. Moreover, the
response of the model is completely determined once the average
and the variance of the stochastic input are given.

In the silent regime we have reported the coexistence of two
coherence resonances: one corresponding to the regularization of
the emitted spike trains at intermediate noise levels and the
other one to the stimulation of subthreshold oscillations
occurring at very low noise. All the employed indicators, namely
the coefficient of variation, the signal and ISI correlation
times, and the conditional entropies, are able to identify the
first resonance, while the second one can be detected only by the
autocorrelation time of the signal. Conditional entropies, used
for the first time in the context of coherence resonance, turns
out to be an effective indicator for coherence. Experimental
evidences of a similar coexistence of two kinds of resonance has
recently been reported in \cite{coexist} for measurements
performed on an electrochemical cell and numerically corroborated
by simulations of a FitzHugh-Nagumo model. However, there are two
differences with our analysis: in Ref.~\cite{coexist} the
subthreshold signal was injected in the system together with the
noise (therefore it is a stochastic resonance, and not a coherence
resonance) and the two resonances are both observable at the level
of spike trains.

The dynamics in the bistable regime can be described by activated
jumping processes across barriers between two stable solutions :
namely, the oscillatory and silent state. The relative stability
of the two dynamical regimes is ruled by the ratio of the
corresponding barrier heights: until the fixed point solution
remains the most stable state the observed dynamics resembles that
in the silent regime. In the repetitive firing regime the only
noticeable feature is related to a maximization of incoherence
observed at a finite noise level.

Moreover, we have clarified the various mechanisms
responsible for spike triggering and for ending repetitive firing.
In particular, at relatively low noise the silent neuron can fire
due to stochastic fluctuations via  two mechanisms: one related to
relaxation oscillations following a spike and another one associated
with noise induced activation processes. The first mechanism
lead to the multi-peaked structure of the ISI distributions while
the latter is responsible for the exponential tail.
The presence of peaks in the $P_{ISI}(t)$ suggests that the system,
under the influence of stochastic inputs, can resonate when forced
with specific frequencies corresponding
to $1/t^{(k)}_{ISI}$, the main peak (due to nonlinear effects)
being associated to frequencies in the $\gamma$-range~\cite{sheperd}
(namely, from $40$ to $66$ Hz for $\bar I \in [0:9] \enskip \mu A/cm^2$),
while the second one to lower frequencies (namely, from $30$ to $37$ Hz
for the same interval of currents). Indeed these results can represent
an explanation of recent findings~\cite{liu_feng_2003},
where clear stochastic resonance effects were observed in the range from $30$ to $65$ Hz
for a Hodgkin-Huxley neuron subjected to Poisson distributed trains of EPSPs and IPSPs
with periodically modulated rates plus a subthreshold harmonic signal.
The authors~\cite{liu_feng_2003}
suggest that the origin of such SR should be related to
subthreshold oscillations, however the corresponding frequencies
(i.e. $61 \le 1/T_l \le 92$ Hz for $\bar I \in [0:9] \enskip \mu A/cm^2$)
are too high to match with the observed resonance.
Once relaxed in the rest state the Hodgkin-Huxley neuron
begins to act as a selective filter responding to current fluctuations
with the same frequency of the linear subthreshold oscillations,
in agreement with the analysis in~\cite{yu_wang_2001}.

In the bistable and repetitive firing regime the multi-modal
structure of $P_{ISI}(t)$ at low noise is related to the periodicity of
suprathreshold and (linear) subthreshold oscillations.
Tonic firing states end when the neuron
is stimulated with fluctuations in anti-phase with
respect to the internal oscillations.
In the high noise limit the neuron dynamics in all the three examined
regimes can be represented as a stochastic Wiener process plus drift.

In conclusion, we have shown that the Hodgkin-Huxley neuron in the
high input regime displays a large variety of dynamical behaviors,
thus rendering the study of its dynamics interesting {\it per se}
and  not only for its biophysical implication. The richness of the
Hodgkin-Huxley dynamics is particularly pronounced in the silent
regime for low input noise variance, where the response of the
model resembles more the activity of cortical neurons, since it is
characterized by an almost Poissonian distribution of interspike
intervals~\cite{shadlen_newsome_1998}. In this regime the single
neuron response can selectively enhance stochastic stimulations
corresponding to different spectral ranges, thus allowing for a
contemporary detection and transduction of a variety of signals.
Moreover, a network of such neuronal elements will have the
capability to exhibit coherent and correlated activity over
different time scales (mainly in the $\gamma$ and $\beta$-ranges
~\cite{sheperd}), a property that is believed to be important for
information encoding for cortical
processing~\cite{salinas_seinoski_nature}. Indeed, it has been
found ~\cite{yu_wang_2001} that a globally coupled Hodgkin-Huxley
network subjected to stochastic inputs reveal, for sufficiently
strong synaptic coupling, a collective synchronized rhythmic
firing in a range of $40-60$ Hz, induced via coherence resonance.

As future developments of the present work, we plan to investigate
the role played by correlations among the synaptic inputs in
enhancing or depressing the coherence effects here discussed.

\acknowledgments
We acknowledge useful discussions with V. Beato, B. Lindner,
R. Livi, A. Politi, L. Schimansky-Geier, and R. Zillmer.
One of us (TK) has been supported by the
Marie Curie Individual Intra-European Fellowship "DEAN", project No 011434.

%%%%%%%%%%%%%%%%%%%%%%%%%%%%%%%%%%%%%%%%%%%%%%%%%%%%%%%%%%%%%%%%%%%%%%%%%%%%%%
%       References
%%%%%%%%%%%%%%%%%%%%%%%%%%%%%%%%%%%%%%%%%%%%%%%%%%%%%%%%%%%%%%%%%%%%%%%%%%%%%%


\begin{thebibliography}{9}

\bibitem{tuckwell} H.C Tuckwell, {\em Introduction to theoretical
neurobiology}, (Cambridge University Press, New York, 1988).

\bibitem{koch} C. Koch, {\em Biophysics of computation},
(Oxford University Press, New York, 1999).

\bibitem{ger} W. Gerstner and W. Kistler, {\em Spiking Neuron Models},
(Cambridge University Press, Cambridge, 2002).

\bibitem{hh} A.L. Hodgkin and A.F. Huxley, {\it J. Physiol.(Lond.)}
\textbf{117}, 500 (1952).

\bibitem{hh_devil} C. Meunier and I. Segev,
{\it TRENDS Neurosci.} \textbf{25}, 558 (2002)

\bibitem{shadlen_newsome_1998} M.N. Shadlen and W.T. Newsome,
J.  Neurosci. \textbf{18}, 3870 (1998).

\bibitem{tiesinga_2000} P.H.E. Tiesinga, J.V. Jos\`e, and T.J.
Sejnowski, \pre \textbf{62}, 8413 (2000)

\bibitem{aguera03} B. Ag{\"u}era y Arcas, A.L. Fairhall, and
W. Bialek, {\em Neural Computation},  \textbf{15}, 1715 (2003).

\bibitem{brown_feng_1999} D. Brown, J. Feng and S. Feerick,
\prl \textbf{82}, 4731 (1999).

\bibitem{hasegawa} H. Hasegawa, \pre \textbf{61}, 718 (2000).

\bibitem{gammaitoni} L. Gammaitoni, P. H{\"a}nggi, P. Jung, and
F. Marchesoni, {\it Rev. Mod. Phys.} \textbf{70}, 223 (1998).

\bibitem{gang} H. Gang, T. Ditzinger, C.Z. Ning, and H. Haken,
\prl \textbf{71}, 807 (1993).

\bibitem{pikovsky_kurths_1997} A.S. Pikovsky and J. Kurths,
\prl \textbf{78}, 775 (1997).

\bibitem{lindner_ojalvo_2004} B. Lindner, J. Garc\`ia Ojalvo,
A. Neiman, and L. Schimansky-Geier, {\it Physics Reports} \textbf{392}, 321 (2004).

\bibitem{manjarrez} E. Manjarrez, J.G. Rojas-Piloni, I. M{\'e}ndez,
L. Martinez, D. V{\'e}lez, D. V{\'a}zquez, and A. Flores,
{\it Neurosci. Lett.} \textbf{326}, 93 (2003).

\bibitem{Pakdaman01} K. Pakdaman, S. Tanabe, and T. Shimokawa,
{\it Neural Networks} \textbf{14}, 895 (2001).

\bibitem{Lindner02} B. Lindner, L. Schimansky-Geier, and A. Longtin,
\pre \textbf{66}, 031916 (2002).

\bibitem{Longtin97} A. Longtin, \pre \textbf{55}, 868 (1997).

\bibitem{Gong02} P.L. Gong, J.X. Xu, and S. J. Hu, {\it Soliton
and Fractals} \textbf{13}, 885 (2002).

\bibitem{lee_neiman_1998} S.G Lee, A. Neiman, and S. Kim,
\pre \textbf{57}, 3292 (1998).

\bibitem{yu_wang_2001} Y. Yu, W. Wang, J. Wang, and F. Liu,
\pre \textbf{63}, 21907 (2001).

\bibitem{neiman_1996} A. Neiman, B. Shulgin, V. Anishchenko, W. Ebeling,
L. Schimansky-Geier, and J. Freund, \prl \textbf{76}, 4299 (1996).

\bibitem{noble&stein} D. Noble and R.B. Stein,
J. Physiol. \textbf{187}, 129 (1966).

\bibitem{rinzel_miller_1980} J. Rinzel and R. Miller,
{\it Math. Biosci.} \textbf{49}, 27 (1980).

\bibitem{bialek} F. Rieke, D. Warland, R. de Ruyter van Steveninck and W. Bialek,
{\em Spikes: Exploring the neural code} (Massachusetts
Institute of Technology: Cambridge, Massachusetts, 1996).

\bibitem{shannon} C.E. Shannon,
Bell Syst. Tech. J., \textbf {27}, (1948) 379-423; {\it ibidem} 623-656.

\bibitem{cover} T.M. Cover and J.A. Thomas, {\it Elements of information
theory} (John Wiley and Sons, New York, 1991).

\bibitem{farmer} J.D. Farmer, {\it Naturforsch.} \textbf{37a}, 1304 (1982).

\bibitem{grass_estimator} P. Grassberger {\it Entropy
estimates from insufficient samplings}, preprint (2003)
physics/0307138.

\bibitem{invgau} R.S. Chikara and J.L. Folks, {\em The inverse
Gaussian distribution} (Marcel Dekker, New York, 1988)

\bibitem{wilbur_rinzel_1983} W.J. Wilbur and J. Rinzel,
J. Theor. Biol. \textbf{105}, 345 (1983).

\bibitem{kramers} H.A. Kramers,
{\it Physica} \textbf{7}, 284 (1940).

\bibitem{coexist} G.J. Escalera Santos, M. Rivera, and P. Parmananda,
\prl {\bf 92} (2004) 230601.

\bibitem{sheperd} {\it The synaptic organization of the brain},
ed. G.M. Sheperd (Oxford University Press, New York, 2004)

\bibitem{liu_feng_2003} F. Liu, J. Feng, and W. Wang,
{\it Europhys. Lett.} {\bf 64}, 131 (2003).

\bibitem{salinas_seinoski_nature} E. Salinas and T.J. Sejnowski,
{\it Nature Reviews in Neuroscience} \textbf{2}, 539 (2001).

\end{thebibliography}
\end{document}